\documentclass[10pt]{paper}
\usepackage{amsmath, amsthm, amssymb, }
\usepackage{float,epsfig}
\usepackage{color}
\usepackage{graphicx}
\usepackage{mathtools}
\usepackage{subcaption}
\usepackage{multirow}
\RequirePackage{graphicx,epstopdf,adjustbox}

\begin{document}

\newtheorem{theorem}{\bf Theorem}[section]
\newtheorem{proposition}[theorem]{\bf Proposition}
\newtheorem{definition}[theorem]{\bf Definition}
\newtheorem{corollary}[theorem]{\bf Corollary}
\newtheorem{remark}[theorem]{\bf Remark}
\newtheorem{lemma}[theorem]{\bf Lemma}
\newcommand{\nrm}[1]{|\!|\!| {#1} |\!|\!|}

\newcommand{\ba}{\begin{array}}
\newcommand{\ea}{\end{array}}
\newcommand{\von}{\vskip 1ex}
\newcommand{\vone}{\vskip 2ex}
\newcommand{\vtwo}{\vskip 4ex}
\newcommand{\dm}[1]{ {\displaystyle{#1} } }

\newcommand{\be}{\begin{equation}}
\newcommand{\ee}{\end{equation}}
\newcommand{\beano}{\begin{eqnarray*}}
\newcommand{\eeano}{\end{eqnarray*}}
\newcommand{\inp}[2]{\langle {#1} ,\,{#2} \rangle}
\def\bmatrix#1{\left[ \begin{matrix} #1 \end{matrix} \right]}
\def \noin{\noindent}
\newcommand{\evenindex}{\Pi_e}

\def\bmatrix#1{\left[ \begin{matrix} #1 \end{matrix} \right]}
\def \noin{\noindent}



\def \R{{\mathbb R}}
\def \C{{\mathcal{C}}}
\def \X{{\mathcal{X}}}
\def \F{{\mathbb F}}
\def \cp{\mathcal{CP}}
\def \cn{{\mathcal{CN}}}
\def \T{{\mathcal{T}}}
\def \NT{{\mathcal{NT}}}
\def \lam{{\lambda}}


\title{A model for the spread of an epidemic from local to global: A case study of COVID-19 in India}
\author{Buddhananda Banerjee\thanks{ Department of Mathematics and Centre for Excellence in Artificial Intelligence, IIT Kharagpur, E-mail: bbanerjee@maths.iitkgp.ac.in},\, Pradumn Kumar Pandey\thanks{Department of Computer Science and Engineering, IIT Roorkee, India Email: pradumn.pandey@cs.iitr.ac.in},\, and  Bibhas Adhikari\thanks{Corresponding author, Department of Mathematics and Center for Theoretical Studies, IIT Kharagpur, India, E-mail: bibhas@maths.iitkgp.ac.in }}

\date{}

\maketitle
\thispagestyle{empty}

\noindent\textbf{Abstract.} In this paper we propose an epidemiological model for the spread of COVID-19. The  dynamics of the spread is based on four fundamental categories of people in a population: Tested and infected, Non-Tested but infected, Tested but not infected, and non-Tested and not infected. The model is based on two levels of dynamics of spread in the population: at local level and at the global level. The local level growth is described with data and parameters which include testing statistics for COVID-19, preventive measures such as nationwide lockdown, and the migration of people across neighboring locations. In the context of India, the local locations are considered as districts and migration or traffic flow across districts are defined by normalized edge weight of the metapopulation network of districts which are infected with COVID-19. Based on this local growth, state level predictions for number of people tested with COVID-19 positive are made. Further, considering the local locations as states, prediction is made for  the country level. The values of the model parameters are determined using grid search and minimizing an error function while training the model with real data. The predictions are made based on the present statistics of testing, and certain linear and log-linear growth of testing at state and country level. Finally, it is shown that the spread can be contained if number of testing can be increased linearly or log-linearly by certain factors along with the preventive measures in near future. This is also necessary to  prevent the sharp  growth in the count of infected and to get rid of the second wave of pandemic.    

\section{Introduction}

COVID-19 is a pandemic that is actively spreading in the
whole world and is an unprecedented challenge for the
human race. All the countries infected with COVID-19 are struggling to mitigate the spread through various strategies. This disease is spread by inhalation or contact with infected droplets or fomites. It is observed that successful medical testing and as a result, detection of people infected with SARs-Cov-2 becomes one of the crucial control strategies for the spread of COVID-19 \cite{salathe2020covid}. For instance, the epidemic curve in The Republic of South Korea suggests that this control strategy in South Korea has curtailed the epidemic. Besides, testing is also linked to tracing contact lists of the infected people and finally self-isolation of those people help against the spread. The success of containment of COVID-19 in the Republic of Taiwan has also the influence of proactive testing \cite{wang2020response}. 

Given the fact that there in no effective antiviral vaccine or drug should coming soon, different prevention strategies are adapted by different countries that include  voluntary or compulsory quarantine, stopping of mass gatherings, closure of educational institutions or workplaces, social distancing or even nationwide lockdown. However, these strategies may act less significant for the infected people who are at the pre-symptomatic stage, and in that case they act as invisible spreaders for the disease \cite{meidan2020alternating}. Thus it becomes increasingly important for mass medical testing for a country. Several researchers around the world are actively working on producing mathematical models of the spread of COVID-19. Here we quote that `model-based predictions can help policy makers make the right decisions in a timely way, even with the uncertainties about COVID-19 \cite{anderson2020will}.

 The primary preventive steps adapted by the Government of India fall into five categories which include social distancing, movement restrictions public health measures, social and economic measures, and nationwide lockdown. A few notable decisions by the Government of India are given in Table \ref{table:measures}. It should be noted that a complete nationwide lockdown from March 25 till May 13 helped to control the spread the disease at large distances but failed to prevent it in neighboring districts, as observed in \cite{pandey2020lockdown}. For example, before lockdown, infected cases are reported from different districts across India which are at large distances apart, however during the lockdown period it has been observed that new spread is reported in districts which are neighbors of infected districts. Besides, due to lack of well planned policy for migrant workers several of them have been travelling to their native districts during lockdown. Unavailability of data of such a traffic flow across the districts can be crucial in order to do a precise analysis of the spread. It can also be seen that the preventive measures proposed by the Government of India are similar to those adapted in other counties. 


It is observed in various studies that COVID-19 exhibits  significantly different epidemiological attributes than other well studied epidemics in past. Thus it is of paramount interest to develop mathematical models which can characterize the inherent dynamics of the spread of COVID-19. Standard epidemic models such as SIR model considers human-to-human transmission, and it describes the diffusion process through three mutually exclusive stages of infection: Susceptible, Infected and Recovered. These models are also called compartmental models \cite{balabdaoui2020age} which enables to compartmentalize different individuals based their states for the epidemic in a population. This model can help gain some insights about the growth of the infection based on approximating the model parameters from the available data. However due to a peculiar growth of COVID-19 in different countries, researchers have extended the SIR model and other existing models such as SIS model in order to acquire  meaningful insights about spread of the COVID-19 \cite{kohler2020robust}. It is very important to note that these studies can help us to frame control strategies and policies that can mitigate the epidemic \cite{rawson2020and} \cite{agosto2020statistical}. 

One of the first models for the spread of COVID-19 is proposed by Anastassopoulou et al. based on the data of confirmed cases reported at the Hubei province of China from the 11th of January until the 10th of February, 2020 \cite{anastassopoulou2020data}. They propose a discrete SIRD (Susceptible-Infected-Recovered-Dead) model and estimate the mean values of the corresponding epidemiological parameters such as basic reproduction number, the case fatality and case recovery ratio from the data. This model enables to forecast about the spread in near future. In an another attempt, in \cite{kucharski2020early} the authors study the datasets of transmission from within and outside Wuhan, China to estimate how transmission in Wuhan varied between December 2019, and February 2020, and assess the potential for sustained human-to-human transmission to occur in locations outside Wuhan through a stochastic transmission dynamic model. In  \cite{giordano2020modelling}, a mean-field epidemiological model is proposed for COVID-19 epidemic in Italy by extending the classical SIR model. Here, in addition to susceptible (S) and infected
(I), the other stages of individuals are considered as diagnosed (D), ailing (A), recognized (R), threatened (T),
healed (H) and extinct (E), collectively termed as SIDARTHE. In \cite{balabdaoui2020age}, an Age-stratified model of the COVID-19 is proposed to capture the age-dependent dynamics for nowcasting and forecasting for Switzerland. This model incorporates the compartments of symptomatic and asymptomatic infected individuals along with susceptible and exposed individuals. In \cite{dziugys2020simplified}, the authors propose a model of COVID-19 epidemic dynamics under quarantine conditions. They also develop methods to estimate quarantine effectiveness in a country or a region which is infected with COVID-19. Besides, a few models are proposed for understanding and predicting the spread of COVID-19 based on metapopulation network approach, see \cite{arenas2020mathematical} \cite{costa2020metapopulation} \cite{wells2020covid}. 

Several mathematical models are also proposed based on the the available COVID-19 data of India and fitting them into classical epidemic models incorporating other factors such as nationwide lockdown, social distancing etc., see \cite{pandey2020seir} \cite{ranjan2020predictions} \cite{dhanwant2020forecasting} and the references therein. In \cite{singh2020age}, a mathematical model of the
spread of COVID-19 is proposed based on an age-structured SIR model. However, the comparison of this model prediction with real data is criticized by Dhar in \cite{dhar2020critique}. In \cite{ghosh2020covid}, the authors perform state-wise analysis of the data of infected population in different states based three models: Exponential Model, Logistic Model and the SIS model. They also provide state-wise prediction for number of infected people for different states in recent future. An elementary network-based model for geographical spread of COVID-19 in India is proposed in \cite{kumar2020modeling} . In \cite{pujari2020multi}, a model for the spread of COVID-19 in India is proposed emphasizing on migration of population based on the spatial network of cities, incorporating the growth-dynamics of SIR model at the city-level.  

In this paper, we propose an epidemiological model for the spread of a contagious epidemic in a region or country. The entire model is based on combining two growth processes of the spread at local and global level. By local, we mean at the level of city or town or districts or province, and global mean at the level of state or country. First we develop a new discrete model for the growth-dynamics of infected people at local level as follows. We consider four type of individuals living at a location. These are individuals who are tested as infected ($X_1$), tested as non-infected ($X_2$), untested but infected (asymptomatic or pre-symptomatic, $X_3$), and untested and non-infected $(X_4)$ for the disease. Total number of such individuals equals the total population living at that location. Given the time series data of these numbers $X_i(t), t=1,2,3,4$, we define the growth statistic $X_i(t+1)-X_i(t)$ utilizing $X_j(t), j\neq i$ and four other parameters each one of them is related to the the spreading pattern of the virus which causes the disease. Note that the different standard compartmental models exist in literature based on susceptible, infected, recovered, and diseased, which do not preserve the effect of parameters in an epidemic like COVID-19. In our proposed model, the growth-dynamics at local level include the following parameters: \begin{itemize}
    \item[(a)] Spread due to infected but asymptomatic and pre-symptomatic individuals 
    \item[(b)] Effect of preventative measures like lockdown or restricted movement of individuals across locations
    \item[(c)] Daily testing statistics.
\end{itemize}  

Then we consider the metapopulation network of all the locations at local level in order to incorporate the transmission dynamics of disease at the global level. Here we mention that the metapopulation network model is a standard and popular model for analyzing the spread of highly contagious diseases which include Zika virus \cite{zhang2017spread}. Also see \cite{wang2014spatial} and the references therein. In our proposed model,  the vertices of the metapopulation network are the locations infected with the disease and the links connecting them represent the possible mode of transportation or spatial distance such as the great circle distance of the latitude and longitude coordinates of the locations at local level. The weight of these links, that represent the rate or percentage of transmission of population per unit time such as a day. Then the final model is defined by combining the dynamics of the spread at local and global level. The values of the model parameters are obtained by a learning technique based on training data and an error minimization.

In the case of COVID-19, we consider the model parameters at the local level as testing statistic, social distance, and rate of infected people by an infected but untested individual (asymptomatic or pre-sympotatic) per unit time. 
In the context of India, the locations are considered as districts which constitutes the states and union territories of India.  There are 28 states and 8 union territories in India, and there are a total of 718 districts in India. 

Based on the proposed model we predict number of COVID-19 infected people both at state level and the country (India) level. The prediction depends on the number of testing performed per day. The results show that the total number of infected people at India level will be approximately 0.46 Millions on July 7, 2020,  1.9 Millions on November 7, 2020, and 4.6 Millions on May 7 2021 when the number of testing is approximately 1,00,000 per day at the country level (which is the number of testing as on May 7, 2020 approximately). If the number of testing grows linearly (with a certain rate see Section \ref{sec:predictionIndia}) then the number of people tested positively with COVID-19 would be approximately 2 Millions on July 7, 2020, 59 Millions on November 7, 2020, and 130 Millions on May 7, 2021. Finally, if the number of testing grows log-linearly (with a certain rate, see Section \ref{sec:predictionIndia}) then the number COVID-19 infected people in India would be approximately  1.3 Millions on July 7, 2020, 3.77 Millions on November 7, 2020, 8.5 Millions on May 7, 2021. Note that these above mentioned predictions are made when there is no external measure is used to control the spread, for example, using any cure like a vaccine or drug discovered in between. Further using numerical simulation we show that the spread stops when daily number of testing increases linearly or log-linearly, however if the number of testing remains approximately the same as of May 7, 2020 the spread need not stop in recent future, say in the year 2021.

\section{The proposed model}

Let $\mathcal{V}=\{~l~ |~ l ~\mbox{is the index of a location}\}$ be the set of locations where persons infected with COVID-19 are likely to stay in or move to on a day $t$.
Suppose that $N_l$ is the population size in location $l$. Now we introduce the following notations to model the distribution and dynamics of pandemic. If $T_l(t)$ denotes the number of tested individuals in the location $l$  then $\bar T_l(t)=N_l-T_l(t)$ stands for the number non-tested individuals   up to time $t$.  Let $C^+_l(t)$ and $C^-_l(t)$  be  the total number of people infected   and non-infected with COVID-19, respectively  in a  location  $l \in \mathcal{V} $.
Here   these temporal data varies  with time $(t)$ measured in days.
In any location $l$ for a given day $t,$ we define a  random vector 
$$\mathbf X^{[l]}(t)=\bmatrix{X_{1}^{[l]}(t) & X_{2}^{[l]}(t) & X_{3}^{[l]}(t) & X_{4}^{[l]}(t)}^T$$
with four components for the  distribution of population the $N_l.$ 
Based on the above discussion $\mathbf X^{[l]}(t)$ can be represented in a $2\times2$ contingency-table, Table \ref{pop_distribution}. Obviously, $$\sum_{j=1}^4 X_{j}^{[l]}(t) = N_l, \forall~ l~\in \mathcal{V}, $$ the total population at the location $l$, though $X_{3}^{[l]}(t) ~ \& ~ X_{4}^{[l]}(t)$ are unobserved or latent  random variables. 
\begin{table}[t]
\begin{center}
\begin{tabular}{ |c|c c|c| } 
\hline
 & COVID$+ve$ & COVID$-ve$& Total \\
\hline
&&&\\
Tested & $X_{1}^{[l]}(t)$ & $X_{2}^{[l]}(t)$ & $T_l(t)$\\ 
&&&\\
Non-tested & $ X_{3}^{[l]}(t)$ & $X_{4}^{[l]}(t)$ & $\bar T_l(t)$ \\ 
&&&\\
\hline
Total & $C^+_l(t)$ & $C^-_l(t)$ &  $N_l$\\
\hline 
\end{tabular}
\caption{Distribution of population in location $l$ at time $t$}
\label{pop_distribution}
\end{center}
\end{table}
Unlike the standard epidemic models, the asymptomatic infected people or  who are infected with COVID-19 but not tested, that is, $X_{3}^{[l]}(t)$ may influence the number  $X_{1}^{[l]}(t')$ at a future date $t'>t.$ Besides, $C^+_l(t)$ highly depends on the contact networks of $C^+_l(t'')$ at a previous date $t'' < t.$ But only $X_{1}^{[l]}(t)$ is observed.   Thus the number of people who are tested for COVID-19 at a given day governs the dynamics of $\mathbf X^{[l]}(t)$ at a location $l$ over time.   

Let $\widetilde{T}_l(t+1)$ be a strategic number  which provides  the target quantity  of new tests for COVID19 to be performed on day $(t+1)$ in location $l.$  Given the statistic $\mathbf X^{[l]}(t),$ new tests also depends on the availability of test-kits. However, this also depends on $\bar T_l(t)$, the number of people not tested for the disease at the location $l.$ Hence, we define the possible  number of tests to be performed at $l$ as 

\be\label{opt:test} \widetilde{T}_l^*(t+1)=\min\{\widetilde{T}_l(t+1),~ \bar T_l(t)\}.\ee

In Table \ref{tab:parameters}, we introduce some generic notations of  model-parameters that are used to  develop the dynamics of the system and some more hyper-parameters that are involved in training and updating of model-parameters. All the  parameters modified with suffix/super-fix according to the time  and locations accordingly.

\begin{table}[h]
    \centering
    \begin{tabular}{c|l}
    \hline
       Parameters  & Interpretations \\
       \hline
        $\lambda_1$ & Testing-coverage probability among the infected \\
        $\lambda_2$ & Infection spreading probability \\
        $\lambda_3$ & Probability of population migration among locations \\
        $\alpha$ & Average family size \\
        $\theta$ & Mobility of individuals\\ 
        $\epsilon$ & Error parameter \\
        \hline
        Hyper-parameters  & Interpretations \\
        \hline 
        $\alpha_1$& Changing rate of $\lambda_1$\\ 
        $\alpha_2$& Changing rate of $\lambda_1$ for future\\ 
        $\beta_1$& Changing rate of $\lambda_2$\\
        $r_1$ &Rate of increment in testing  under linear growth. \\
         $r_2$ & Rate of increment in testing  under log-linear growth. \\
         
        \hline
    \end{tabular}
    \caption{Model parameters and hyper-parameters }
    \label{tab:parameters}
\end{table}

Now we define the  dynamics of change of $\mathbf X^{[l]}(t)$ for any location $l.$ 
\begin{eqnarray}
X_{1}^{[l]}(t+1)- X_{1}^{[l]}(t) =\Delta_t X_{1}^{[l]} &=& \mbox{bin}\left(\min\{\widetilde{T}_l^*(t+1), X_{3}^{[l]}(t)\}, \, \lam_1^{[l]}(t+1)\right) \label{eqn:mod1}\\
X_{2}^{[l]}(t+1)-X_{2}^{[l]}(t) =\Delta_t X_{2}^{[l]} &=& \min\{\widetilde{T}_l^*(t+1)-\Delta_t X_{1}^{[l]}, X_{4}^{[l]}(t)\}, \\
X_{3}^{[l]}(t+1)-X_{3}^{[l]}(t) =\Delta_t X_{3}^{[l]} &=&\max\{ - \Delta_t X_{1}^{[l]} +\min\{ a^{[l]}(t+1), X_{4}^{[l]}(t)\}, -X_{3}^{[l]}(t) \}\\
X_{4}^{[l]}(t+1)-X_{4}^{[l]}(t) =\Delta_t X_{4}^{[l]} &=&\max\{ - \Delta_t X_{2}^{[l]} -\min\{ a^{[l]}(t+1), X_{4}^{[l]}(t)\}, -X_{4}^{[l]}(t) \} \label{eqn:mod4}
\end{eqnarray} where
\beano a^{[l]}(t+1) &=& \mbox{bin}(\alpha \Delta_t X_{1}^{[l]}, \lam_2^{[l]}(t+1)) + \mbox{Pois} \left(\lam_3^{[l]}(t+1)\sum_{k=1}^N m_{kl}(t) X_{3}^{[l]}(t)\right) + \mbox{Pois}(\epsilon).\eeano 

$\lam_1^{[l]}(t+1) \in (0,1)$ is testing-coverage probability among the infected in location $l$ at time $(t+1)$. Hence, only a fraction  of $X_{3}^{[l]}(t)$ will be will be identified as  $\Delta_t X_{1}^{[l]}$. So it is modelled with binomial distribution.    $\lam_2^{[l]}(t+1) \in (0,1)$ is a probability indicating the average spread of infection among near by people of a group of infected individuals. So, new spread identified-infected  people is also modelled with binomial random variable  $\mbox{bin}(\alpha \Delta_t X_{1}^{[l]}, \lam_2^{[l]}(t+1))$.  Now $\lam_3^{[l]}(t+1) \in (0,1)$ is a probability closed to zero indicating the influence from adjacent locations. As a consequence it is modelled with $\mbox{Pois} \left(\lam_3^{[l]}(t+1)\sum_{k=1}^N m_{kl}(t) X_{3}^{[l]}(t)\right)$. Parameter  $\epsilon>0$ stands for  average noise with Poisson distribution.  It may be noted that $X_{3}^{[l]}(t)$ and $X_{4}^{[l]}(t)$ are latent variables at a given day $t.$ The parameters $\lam_j(t+1),$ $j=1,2,3$ are to be accessed by a suitable mechanism defined in the next section.

Now we consider the meta-population network $G(t)$ with vertex set $\mathcal{V}$ of locations in order to incorporate the effect of transmission of COVID-19 across the locations. Let $A(t)=[a_{kl}(t)]$ denote the adjacency matrix associated with $G(t).$ Let $d_{kl}$ denote the distance between $k$ and $l.$ Then define the weights of the edges of $G(t)$ as 
\be\label{weight} w_{kl}(t) \propto \exp\left\{- \frac{d_{kl}}{\theta(t)}\right\}\ee where $\theta(t)$ is the mobility parameter. Here $w_{kl}$ denotes diffusion weight for the human traffic flows per day between the neighboring locations $k$ and $l.$ The value of $\theta(t) >0$ may be controlled based on government policies. For instance, in the case of strict lockdown the value of $\theta(t)$ may be considered as a small value.
Now we define the matrix $M(t)=[m_{kl}(t)]$ where $$m_{kl}=\dfrac{w_{kl}(t)}{\sum_{l=1}^{|\mathcal{V}|}w_{kl}(t) }$$ which is a row-stochastic matrix. Finally  we propose the following predictive model at the level of state and country for the number of COVID-19 infected people.


Note that the traffic flow between locations influences the value of $X_4^{[l]}(t+1)$ as followed by Eq.~ (\ref{eqn:mod4})  which contribute to $X_3^{[l]}(t+1)$ and finally to the number of infected people $X_1^{[l]}(t+1).$ Besides, the number of nodes in the metapopulation network $G(t)$ varies with time. At time $t,$ the nodes of $G(t)$ represented the districts which are affected by the diaease at time $t.$ Thus at the level of state $S$ which consists of some locations, $X_1^{[S]}=\sum_{l\in S} X_1^{[l]}$ at anytime $t.$ Further, the number of infected people at the country level is calculated based on the proposed dynamics of $X_i^{[l]}, i\in\{1,2,3,4\}$ where $l$ is a state. This is done presumably due to the traffic flow between neighboring districts may be different from the traffic flow between neighboring states. Hence, at the country level, say India, denoted by $I,$ $X_1^{[I]}=\sum_{S\in I} X_1^{[S]}$ at anytime (day) $t.$


\subsection{ Hyper-parameter selection and model parameters updation} 

In this section we discuss how to determine the values of the parameters involved in the proposed epidemiological model. Note that the initial values can be assumed wisely based on its characteristics observed from data and then as the time passes the model can update  the values of the parameters from observed and simulated data. Let $[t_0, \, t_1]$ be the learning period throughout which the real data is available and the model can learn the data for estimating the values of the parameters. Consequently, the growth-dynamics of parameters can be defined which can update the values of the parameters when the real data is not available in future.    

First we consider the parameter $\lambda^{[l]}_1(t).$ Then define 
\begin{equation}
\lambda^{[l]}_1(t+1)=\lambda^{[l]}_1(t)+\alpha_1 \frac{\widetilde{X}^{[l]}_1(t)-X^{[l]}_1(t)}{m^{[l]}},
    \label{lambda1}
\end{equation}
for some $\alpha_1\geq 0$ and  $$m^{[l]}= \max_{t'\leq t}  |\widetilde{X}^{[l]}_1(t')-X^{[l]}(t')|,$$ where, $\widetilde{X}^{[l]}_1(t)$ is  the reported  number of tested-positive cases in location  $l$ at time $t$, and $X^{[l]}_1(t)$ is the value of tested-positive cases obtained from simulation. In Eq.~(\ref{lambda1}), infection spreading rate $\lambda^{[l]}_1(t)$ is updated in such a way that if the number of tested and infected cases are more than the simulated values then infection spreading rate would be more as compared to current rate of of infection and vice versa. The value of $\alpha_1$ represents the slope of the line along which $\lam_1^{[l]}(t)$ increases with time linearly.

\begin{figure}[ht]
  \centering
\includegraphics[trim = 50 40 0 0, clip,width=0.5\linewidth]{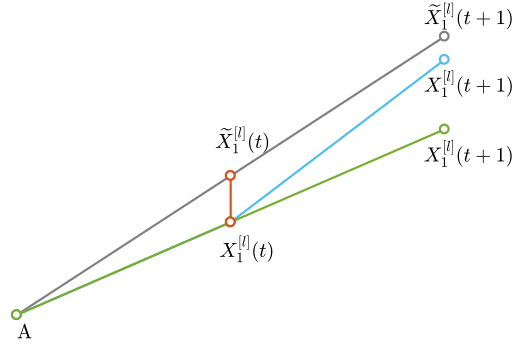}  
\caption{Figure explains the way to update infection spreading rate.}
\label{fig:learning}
\end{figure}

 Eq.~(\ref{lambda1}) is explained pictorially, in Figure~\ref{fig:learning}, consider that points connected via black lines are corresponding to real data points, and points connected via green lines are corresponding to simulated points using $\lambda^{[l]}_1(t)=\lambda^{[l]}_1(t+1)$. In such scenario error $\widetilde{X}_1^{[l]}(t)-X_1^{[l]}(t)$ increases. For better fit of the model we need to update the parameter $\lambda^{[l]}_1(t)$ in such a way that $X^{[l]}_1(t+1)$ can come closer to $X'^{[l]}_1(t+1)$.
In figure, $\widetilde{X}_1^{[l]}(t)>X_1^{[l]}(t)$. $X^{[l]}_1(t+1)$ is the number of tested positive cases, if we increase the rate of infection spread $\lambda^{[l]}_1(t)$ then we can get $X^{[l]}_1(t+1)$
closer to $\widetilde{X}^{[l]}_1(t+1)$, point $X^{[l]}_1(t+1)$ connected to point $X^{[l]}_1(t)$ via blue line.

For the growth of $\lam_2^{[l]}(t)$ over time which represents the probability of the spread of the disease at a location $l.$ Thus we define  
\begin{equation}
    \lambda^{[l]}_2(t+1)= \lambda^{[l]}_2(t)\left(1-\beta_1\frac{\widetilde{T}_l(t+1)-\widetilde{T}_l(t)}{\sum^N_{j=1}\widetilde{T}_j(t+1)}\right),
    \label{lambda2}
\end{equation}
where $\beta_1\geq 0$, $\widetilde{T}_l(t)$ denotes the number of tests performed at the location $l$ at time $t$. Here observe that, the intuition behind Eq.~(\ref{lambda2}) is that the probability of spread of the disease depends on the number of testings done at the location $l.$  We consider constant values of $\lambda_3^{[l]}(t)$ and $\epsilon$ in current version.

\subsection{Testing rate}

\begin{figure}[ht]
  \centering
\includegraphics[trim = 0 0 0 0, clip,width=0.6\linewidth]{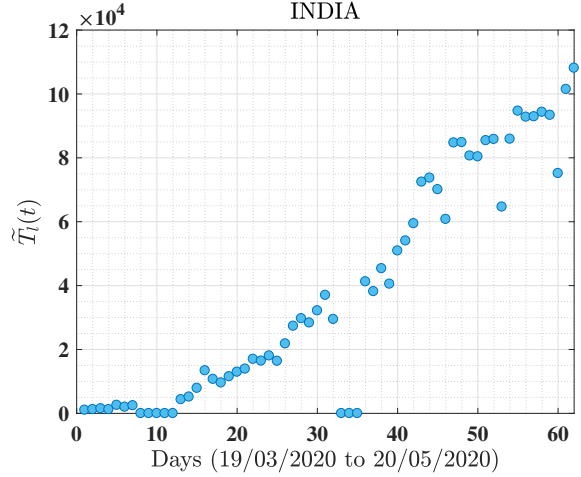}  
\caption{Testing performed daily in India, from March 19,2020 to May 20,2020 \cite{data1}.}
\label{fig:testing}
\end{figure}

 Recall that individuals who are at the asymptomatic and pre-symptomatic stages of infection, act as invisible spreaders for the disease. Hence, detection of individuals who are infected with the virus plays an important role into the growth-dynamics of the number of infected individuals at a particular location. Thus one of the control strategies to prevent the spread is to conduct enough number of tests per day and separate-out the infected people. In a country like India, where approximately 1.4 billion people live, conducting enough tests per day could be a difficult exercise. Besides, due to lack of huge number of test-kits and medical facilities, India is facing a lot of challenges to perform enough tests per day. The testing data in India is plotted in  Fig.~\ref{fig:testing} which is obtained from \cite{data1}. It may be observed that the data is not available for three consecutive days after the 30th day. Besides the testing data is not available before March 19, 2020.
 

 Note that testing for COVID-19 for random sampling of individuals is not desired due to scarcity of enough testing kits for a large population and medical support facilities. Indeed, targeted testing by tracing social contacts of newly detected individuals with COVID-19 can be more efficient for identifying asymptomatic and pre-symptomatic individuals who are infected with the virus. Hence the increment of number of testing per day should depend on the testing-coverage probability among the infected individuals at a particular location, that is, $\lam_1^{[l]}.$
 
 In this model we incorporate two possible growth of testing data over time at a location: linear and log-linear. The parameters which we call rate of gain in the number of tests for COVID-19, are denoted by $r_1$ and $r_2$ for the following linear and log-linear growth equations respectively. From the real data it can be observed that the number of tested positive cases has positive correlation ($0.9177$) with number of test performed. Indeed, from Eq.~(\ref{lambda1}), $\lambda_1^{[l]}(t)$ has positive dependency over the  number of tested positive cases. Thus, the number of tests performed has positive relation with $\lambda_1^{[l]}$.
 
 
Let $\widetilde{T}_{l}(t)$ denote the number of tests performed at a location $l$ on a day $t.$ Then define  

\begin{equation}
 \mbox{linear increment of  testing:} \,\,\,\,   \widetilde{T}_l(t+1)=\left \lceil{\widetilde{T}_l(t)}+r_1\;\lambda^{[l]}_1(t)\right \rceil.
    \label{testing1}
\end{equation}
and
\begin{equation}
  \mbox{log-linear increment of testing:} \,\,\,\,  \widetilde{T}_l(t+1)=\left \lceil{\left(1+r_2\;\lambda^{[l]}_1(t)\right)\widetilde{T}_l(t)}\right \rceil,
    \label{testing}
\end{equation}

Thus assigning small values of $\lambda^{[l]}_1(t), \lambda^{[l]}_2(t)$ in the beginning of the simulation of the model, $\lambda^{[l]}_1(t)$, $\lambda^{[l]}_2(t)$, and $\widetilde{T}_l(t)$ are updated according to Eqs.~(\ref{lambda1}), (\ref{lambda2}), and (\ref{testing}) or (\ref{testing1}) respectively. Further, $\alpha_1$, $\beta_1$, and $r_2$ can be selected from the interior of the unit cube  given by $(0, 1)\times (0, 1) \times (0, 1),$ whereas  $r_1$ can be larger than 1. The searching method  is well-known as as three dimensional \textit{grid search}. Indeed, mapping the growth given by Eqs.~(\ref{lambda1}), (\ref{lambda2}), and (\ref{testing}) with real data, the values of $\alpha_1$, $\lambda^{[l]}_1(t)$, $\beta_1$, $\lambda^{[l]}_2(t)$, $r_2$ (or $r_1$), and $\widetilde{T}_l(t)$ can be learned and estimated such that the total testing $\left(\sum_l\widetilde{T}_l(t')\right)$, and total tested and infected cases $\left(\sum_lX_1^{[l]}(t')\right)$ at time $t'\leq t$ that are close to real data. It is discussed in details in the next subsection.  These estimated values can be used for the training of the model. 

Suppose that $\alpha_2$, $\lambda^{[l]}_1(t')$, $\beta_2$, $\lambda^{[l]}_2(t')$, $r_2'$ (or $r_1'$), and $\widetilde{T}_l(t')$ are the learned values from the given data.
However, for any $t > t'$ when the real data are not available, the trained model can be used for prediction. Thus we define the update of $\lam_1^{[l]}(t)$ as follows:

\begin{equation}
\lambda^{[l]}_1(t+1)=\lambda^{[l]}_1(t)+\alpha_2 \frac{X_1^{[l]}(t)-2X_1^{[l]}(t-1)+X_1^{[l]}(t-2)}{\sum_{l=1}^N\left(X_1^{[l]}(t)-X_1^{[l]}(t-1)\right)},
    \label{lambda11}
\end{equation}



\subsection{Model accuracy measure}\label{modelaccuracy}
Let $X_1^{[l]}(t)$ and $\widetilde{X}^{[l]}_1(t)$ be  the simulated and observed  numbers of detected after test as infected with COVID-19 respectively at a location $l$ at the time (day) $t.$ Consider the time series of real data $\widetilde{X}^{[l]}_1(t)$ where $t_0 \leq t\leq t_1,$ for a particular location $l\in \mathcal{V}$ which is the vertex set of the metapopulation network. Then the complete observed  data-set is given by $\mathbf{X}_1=\{\widetilde{X}_1^{[l]}(t) : l\in \mathcal{V}, t_0\leq t\leq t_1\}.$ Then the data $\mathbf{X}_1$ is divided into two sets which we call the training set and validation set as follows for estimating the model parameters which define $X_1^{[l]}(t).$ Let $t'\in (t_0, \, t_1).$ Define \begin{eqnarray}
X_1^{T} &=& \{\widetilde{X}_1^{[l]}(t) : l\in\mathcal{V}, t\leq t'\} \,\, \mbox{(Training set)}\\
X_1^{V} &=& \{\widetilde{X}_1^{[l]}(t) : l\in\mathcal{V}, t' < t\leq t_1\}. \,\, \mbox{(Validation set)}
\end{eqnarray} The model parameters are calculated which minimize the error function \begin{equation}
    e = wT_e + (1-w)V_e
\end{equation} where \begin{eqnarray}
w &=& \frac{|X_1^V|}{|X_1^T|+|X_1^V|} \\ T_e &=& \frac{1}{|X_1^T|} \sum_{l\in\mathcal{V},\; \widetilde{X}^{[l]}_1(t) \in X_1^T } |\widetilde{X}^{[l]}_1(t)-X^{[l]}_1(t)|\\ V_e
&=&\frac{1}{|X_1^V|} \sum_{l\in\mathcal{V},\; \widetilde{X}^{[l]}_1(t) \in X_1^V } |\widetilde{X}^{[l]}_1(t)-X^{[l]}_1(t)|.
\end{eqnarray} 

Note that the weight $w$ is defined such that $T_e$ and $L_e$ are computed over two different sets $X_1^V$ and $X_1^T$ respectively to avoid the imbalances in the data. 


\subsection{Prediction of number of individuals infected with COVID-19 at global level}

Now we discuss how the values of the model parameters estimated by real data at local location can be used to predict the number of infected people at a global level such as state and country level in near future. We propose to train the model based on two methodologies at the the state level and country level. Recall that a state in India consists of several districts (locations denoted by $l$), and in India there are 28 states and 8 union territories. In this paper we adapt two-step approach for the prediction. The global level parameters include the social mobility parameter $\theta$, and the traffic flow across the local level locations, given by the edge weight of the metapopulation network. 

First, we make state level prediction, that is, $X_1^{[S]}= \sum_{l\in S} X_1^{[l]}(t)$ when $X_1^{[l]}$ is considered at district level $l$, where $S$ is a state of India. The metapopulation network for a state $S$ is formed by the vertices which are districts belong to the state $S,$ and the traffic flow which is represented by weights $w_{kl}$ defined by Eq.~ (\ref{weight}). The distance $d_{kl}$ between two districts $k, l$ is defined by the great circle distance between the longitude and latitude coordinates of $k$ and $l.$ 

Next, once the estimates for $X_1^S$ are obtained for all states $S$ in India, the prediction at the the nation level is obtained by applying the proposed model treating the location as states. Thus model parameters are further estimated comparing with the real data at the level of states, as described above. Further, the metapopulation network of states is constructed, and the traffic flow is calculated using the wight formula $w_{kl}$ where the distance between two states is considered as the great circle distance between the longitude and latitude coordinates of states $k$ and $l.$   


Note that both predictions at state and country level incorporate the social mobility parameter $\theta$ which preserves the effect of policies of the Government. For instance, during locklown the value of $\theta$ is considered as around $50$ (more weights to local travel), and it will take the value around $2000$ (includes long distance travel) when there is no lockdown. Besides the metapopulation network between the locations $l$ plays a crucial role into the prediction. The rate of traffic between two locations is considered as given by Eq. (\ref{weight}). Observed that the effect of social mobility of individuals is also incorporated with the traffic flow. 




\begin{figure}[t]
\begin{center}
\begin{subfigure}{.45\textwidth}
  \centering
  \includegraphics[width=1\linewidth]{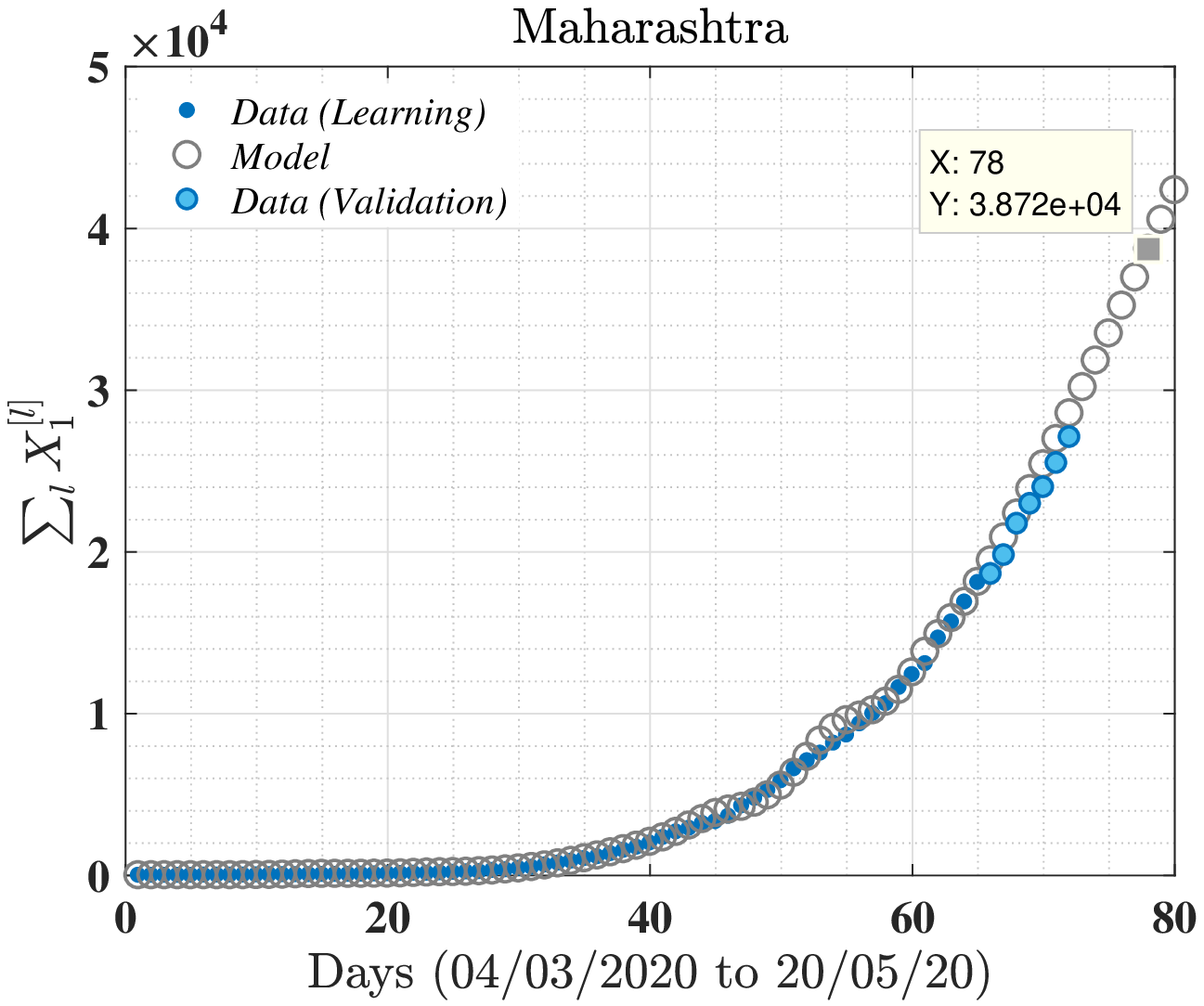}  
  \caption{}
  \label{fig1a}
\end{subfigure}
\begin{subfigure}{.45\textwidth}
  \centering
  \includegraphics[width=1\linewidth]{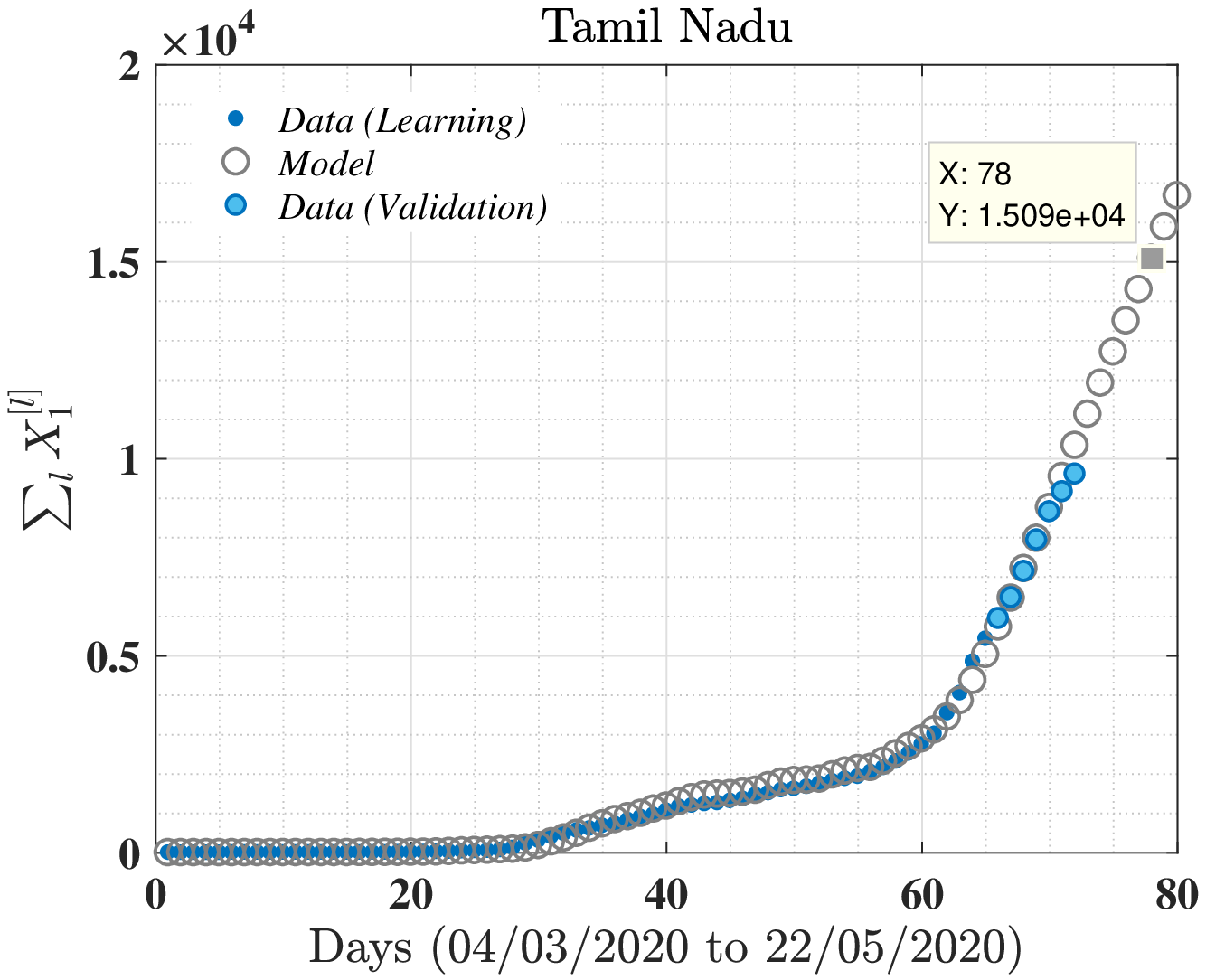}  
  \caption{}
  \label{fig16a}
\end{subfigure}

\begin{subfigure}{.45\textwidth}
  \centering
  \includegraphics[width=1\linewidth]{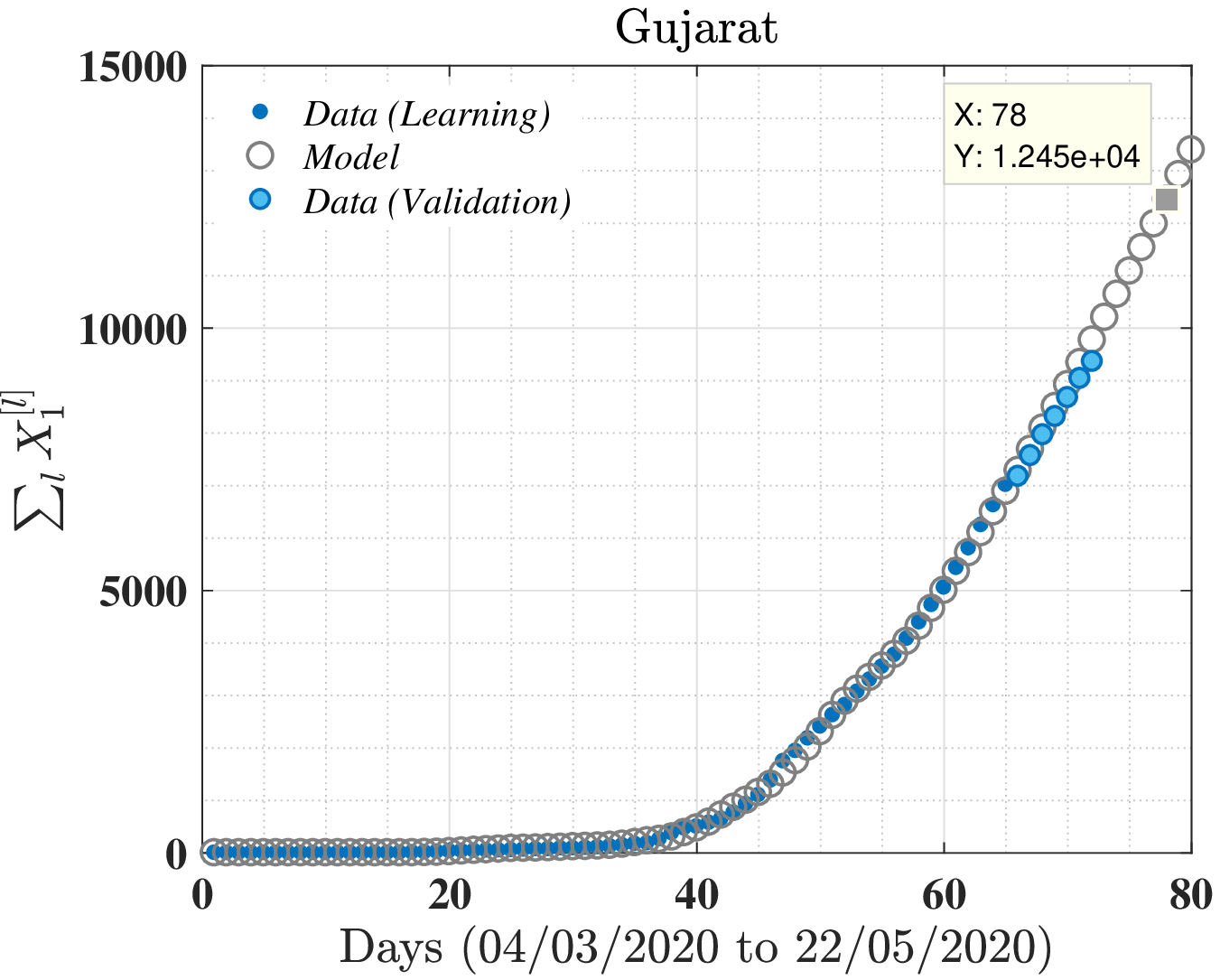}  
  \caption{}
  \label{fig1b}
\end{subfigure}
\begin{subfigure}{.45\textwidth}
  \centering
  \includegraphics[width=1\linewidth]{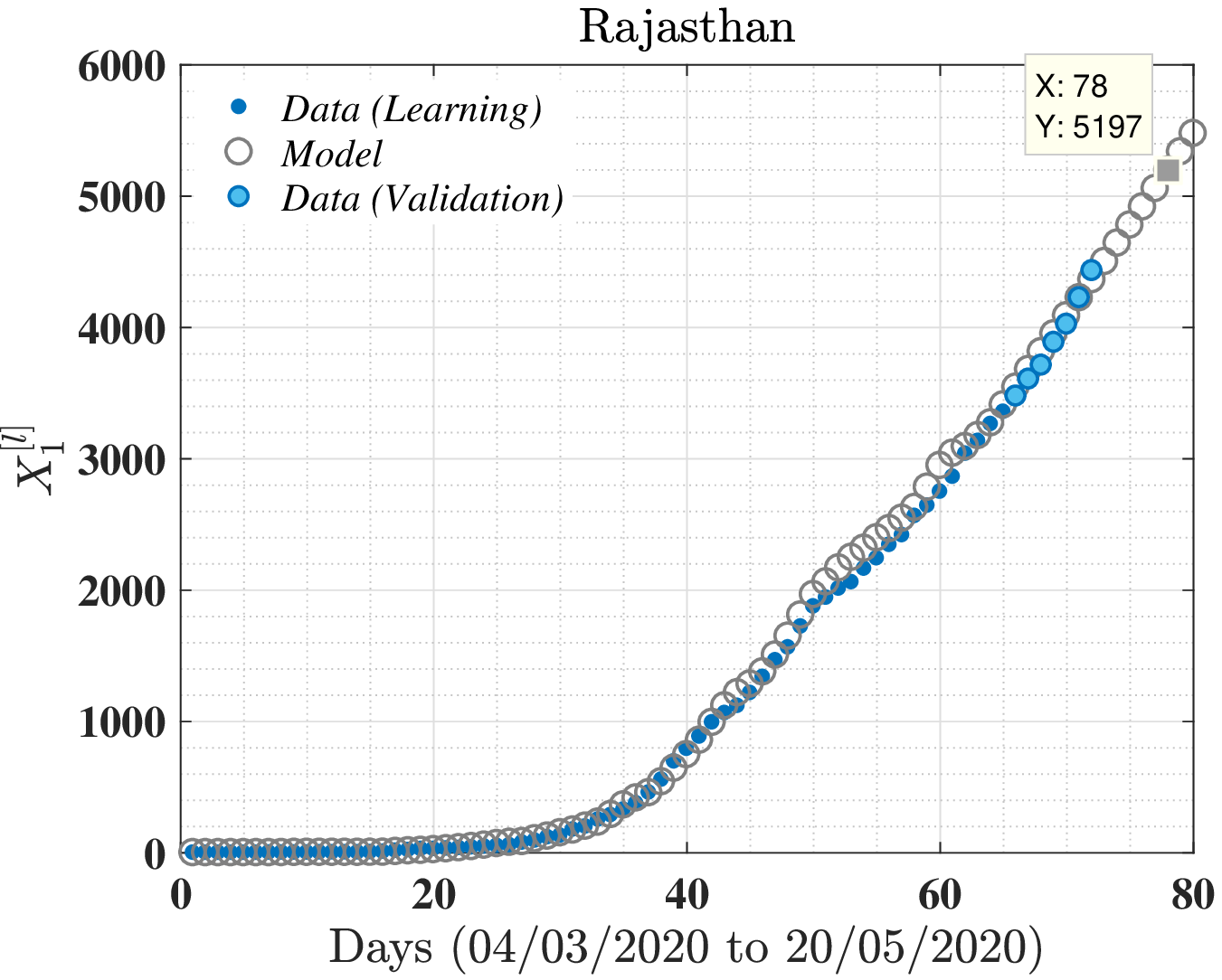}  
  \caption{}
  \label{fig16b}
\end{subfigure}
\end{center}
\caption{Plots are corresponding to data fitting (0-65 days), validation (66-72 days), and prediction of 78th day or May 20, 2020.}
\label{fig:maharashtra}
\end{figure}

\begin{figure}[h]
\begin{center}
\begin{subfigure}{.45\textwidth}
  \centering
  \includegraphics[width=1\linewidth]{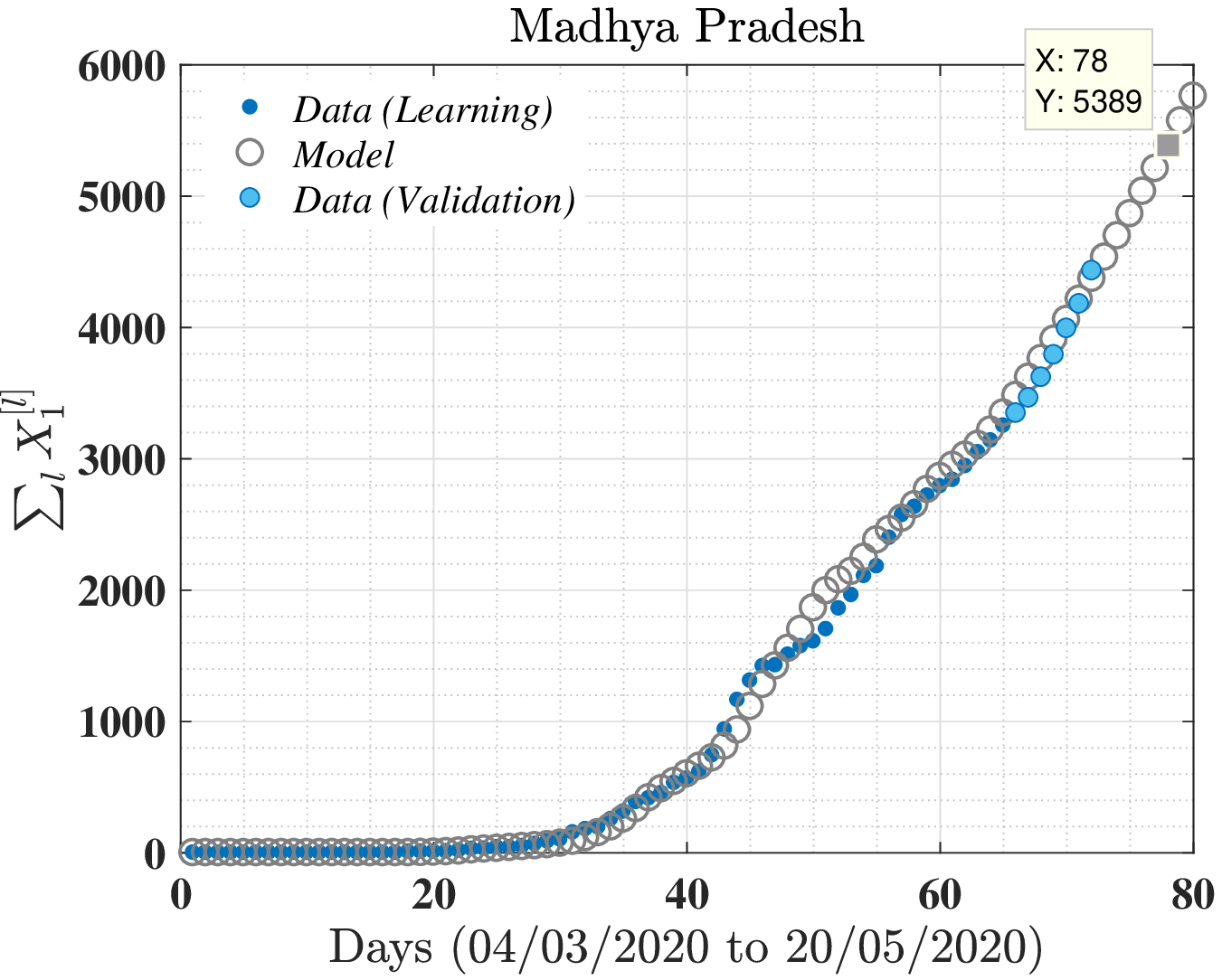}  
  \caption{}
  \label{fig1c}
\end{subfigure}
\begin{subfigure}{.45\textwidth}
  \centering
  \includegraphics[width=1\linewidth]{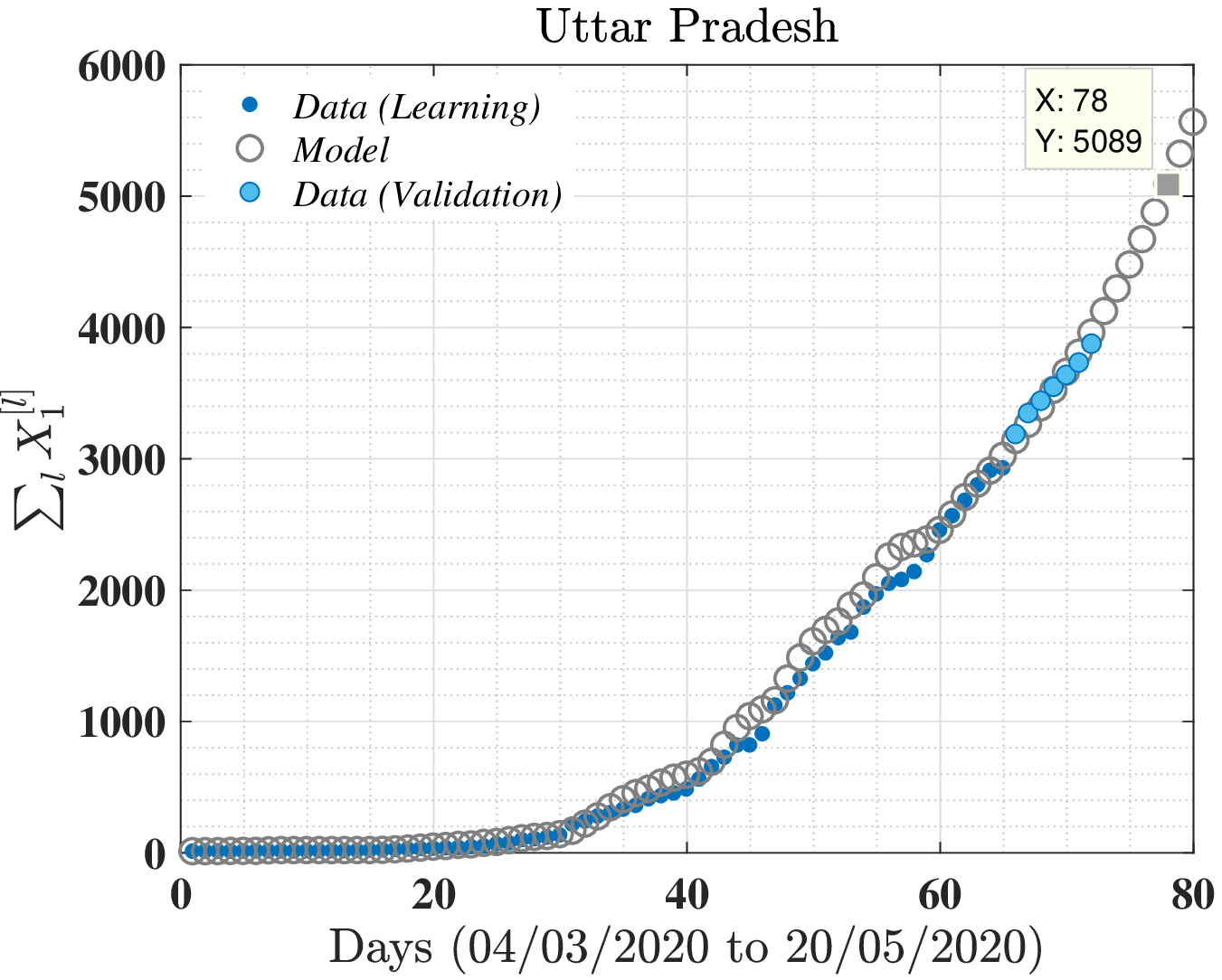}  
  \caption{}
  \label{fig16c}
\end{subfigure}

\begin{subfigure}{.45\textwidth}
  \centering
  \includegraphics[width=1\linewidth]{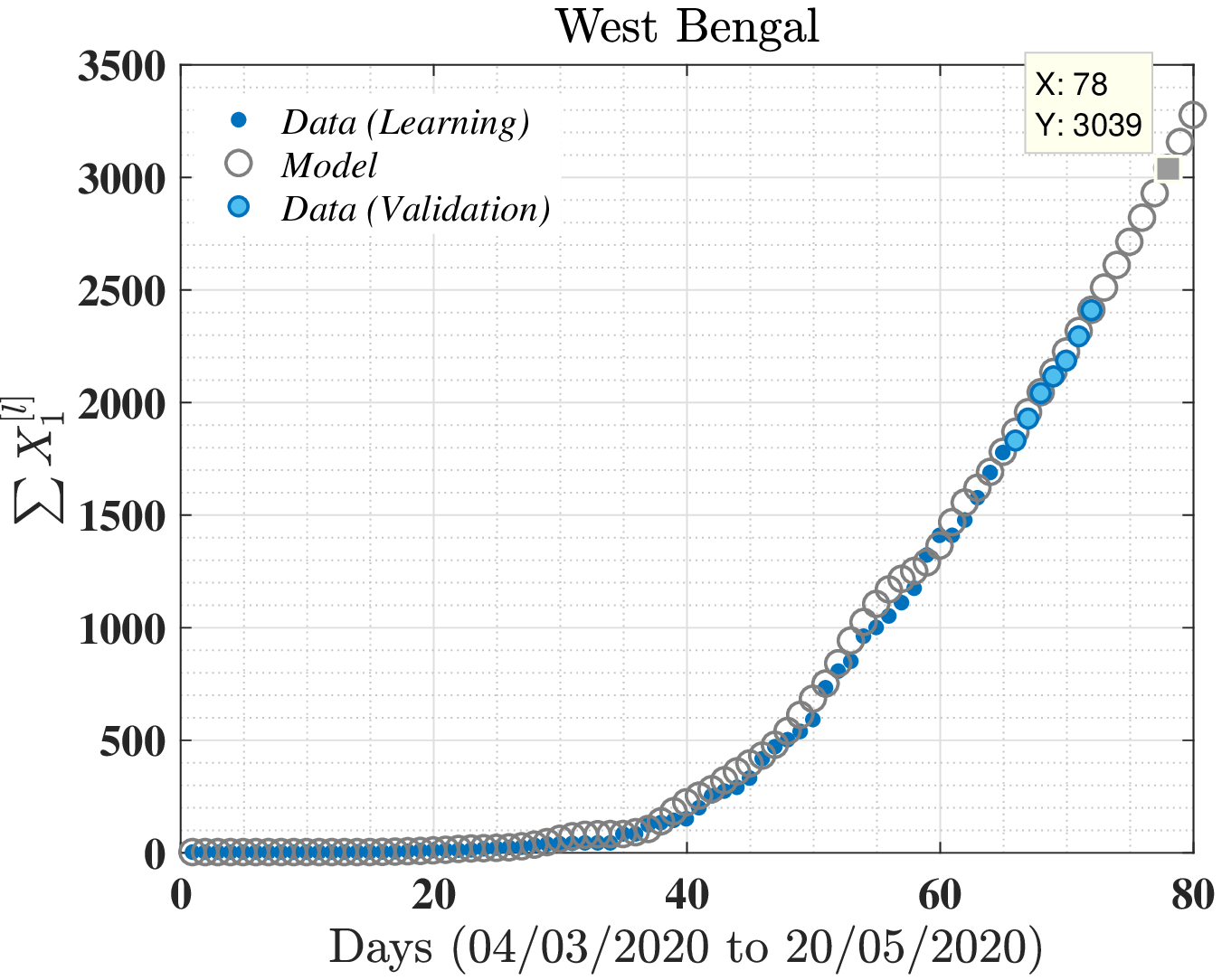}  
  \caption{}
  \label{fig1d}
\end{subfigure}
\begin{subfigure}{.45\textwidth}
  \centering
  \includegraphics[width=1\linewidth]{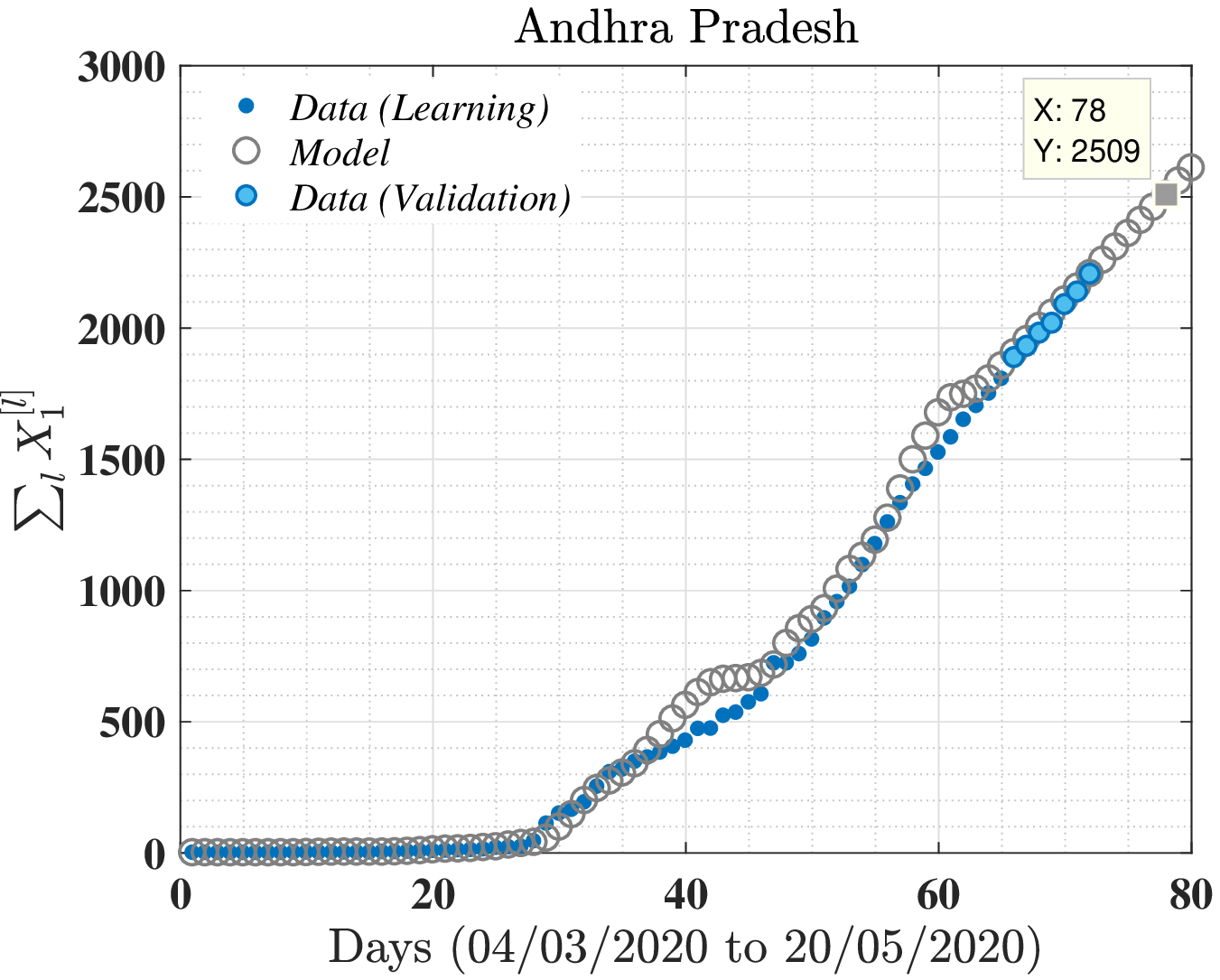}  
  \caption{}
  \label{fig16d}
\end{subfigure}
\end{center}
\caption{Plots are corresponding to data fitting (0-65 days), validation (66-72 days), and prediction of 78th day or May 20, 2020.}
\label{fig:MP}
\end{figure}


\section{Prediction with model and real data: a case study of India}
In this section, the proposed model is trained with the data of infected population with COVID-19 and number of testings performed in India from March 4, 200 to May 7, 2020 \cite{data2}. Since there is nationwide lockdown during this period, the traffic flow across states is less. Therefore we simulate the model at an initial time $t_0=0$ which is on March 4, 2020 by setting $\lambda_1^{[l]}(t_0)=1 \times 10^{-3},$  $\lambda_2^{[l]}(t_0)=3\times 10^{-1}$, $r_2 = 2.8\times 10^{-1}$ and $\lambda_3^{[l]}(t)=1\times 10^{-3}$ for all $t\geq t_0,$ for every location $l.$ These values are assumed due to the following facts. \begin{itemize}
    \item[(1)] Testing-coverage probability is very small since the number of people infected with COVID-19 in the beginning of the spread is small.
    \item[(2)] Since the average family size is 4, so approximately 3 people out of 10 may be exposed to get infected assuming that a person be in close and frequent contact with an group of infected people. 
    \item[(3)]  For simplicity  of the model, we consider the value of $r_2$ such that total number of tested individuals are close to real data. We obtained $r_2=0.28$ during the optimization (training) using grid search.  To match the number of testing performed each day, approximately $1,00,000$ per day given after Eq.~(\ref{testing}).

    \item[(4)] Due to nationwide lockdown, the traffic flow across the locations is less. Hence, $\theta \leq 70$ and $\lambda_3(t)$ is very less. The values of $\theta$ and $\lambda_3(t)^{[l]}$ are obtained using grid search. During error optimization, $(\alpha_1, \beta_1, r_2, \theta, \lambda_3^{[l]})$ is selected from five dimensional grid search.
\end{itemize} 

After the initializing the model the parameter values are learned based on the real data. For instance, the number of testings throughout the period March 04, 2020 to May 20, 2020 in India is approximately $10^5$ per day \cite{data2, data1}, and hence value of $r_2$ is kept fixed during the training period of the model with real data. The value of number of testing at a location $l$ is assumed as a random number between $1$ to $5$ when a first case of COVID-19 is reported. 


The model is trained with the real data collected from How India Lives \cite{data2, data1} for the period of March 04, 2020 to May 07, 2020 (65 days data). The remaining data, that is, the real data for the period May 8 - 20, 2020 is validated based on estimated values of the model parameters of the training dataset. A particular emphasis is given on the states where there are more that 2000 cases of individuals infected with COVID-19 on May 7, 2020 (data used for training) which include Maharashtra (MH), Tamilnadu (TN), Gujrat (GUJ), Rajathan (RAJ), Madhya Pradesh (MP), Uttar Pradesh (UP), West Bengal (WB) and Andhra Pradesh (AP).  The error function contains the absolute difference between data point (observed value) and corresponding value calculated using model. We define an error function which minimizes the absolute mean error of training and validation set, and trained model is used for prediction purpose. The model parameters are learned by
optimizing the error function as described in Section \ref{modelaccuracy}. On  an average per location (district)  error when the prediction is made at the state level is given by 10.7050 (MH), 4.8018 (TN), 5.1519 (GUJ), 4.5936 (RAJ), 2.6368 (MP), 2.3711 (UP), 2.5062 (WB) and 5.2155 (AP) when the number testing grows log-linearly as per the rate $r_2=0.28.$ At the country level the error corresponding to real data and model based prediction is calculated with error 37.2070 on average at each state when the testing grows log-linearly as above. Below we provide prediction at the level of state and India. During training the model, we consider log-linear gain in rate of testing, and error in fitting real data is reported in Table~\ref{table:modelpara}. After training and validation, two cases of gain in rate of testing are considered: linear and log-linear.

\begin{table}[t]
\begin{center}
\resizebox{\textwidth}{!}{ 
\begin{tabular}{ |c|c|c |c|c|c|c|c |c|c| } 
\hline
Parameter & MH &TN&GUJ&RAJ&MP&UP&WB&AP&INDIA\\ \hline
$\alpha_2$ & 0.35 &0.75 &0.45&0.62 &0.53 & 0.61 & 0.23 & 0.51 & 0.49\\
$\beta_1$  & 0.01 &0.005 &0.027&0.11& 0.25 & 0.21 & 0.095 & 0.07 &0.08\\
$\theta$   & 70   &10   &50&70& 70 &50&70&10&70\\
$\lambda_3^{[l]}(t)$ & 1/100 &1/100 & 1/100 & 1/20 & 1/10 & 1/1000 & 1/100 & 1/10 &1/1000\\
$error$ & 10.7050 &4.8018 & 5.1519 & 4.5936 & 2.6368 & 2.3711 & 2.5062 & 5.2155 &37.2070\\
\hline 
\end{tabular}
}
\caption{Trained values Model parameters }
\label{table:modelpara}
\end{center}
\end{table}

\begin{table}[b]
\begin{center}
\resizebox{\textwidth}{!}{ 
\begin{tabular}{ |c|c|c |c|c|c|c|c |c|c| } 
\hline
Day & $r_1$ & MH &TN&GUJ&RAJ&MP&UP&WB&AP\\ \hline \hline
May 20, & $0$ & $3.9\times 10^4$ &$1.3\times 10^4$ &$1.25\times 10^4$ &$6.0\times 10^3$ &$5.7\times 10^3$ & $5.2\times 10^3$ & $3.1\times 10^3$ & $2.56\times 10^3$  \\
2020 & $0$ & $3.87\times 10^4$ &$1.5\times 10^4$ &$1.24\times 10^4$ &$5.2\times 10^3$ &$5.4\times 10^3$ & $5.1\times 10^3$ & $3.0\times 10^3$ & $2.5\times 10^3$  \\ \hline
July 7, & $0$ & $1.3\times 10^5$ &$5.1\times 10^4$ &$4.4\times 10^4$ &$1.2\times 10^4$ &$1.3\times 10^4$ &$9.0\times 10^3$ &$6.0\times 10^3$ &$4.7\times 10^3$ \\
2020  & $10^3$ & $1.6\times 10^7$ &$7.6\times 10^6$ &$2.0\times 10^6$ &$1.5\times 10^6$ &$4.1\times 10^5$ &$2.5\times 10^5$ &$3.7\times 10^4$ &$9.9\times 10^4$ \\
(60 days) & $5\times 10^3$ & $2.1\times 10^7$ & $1.77\times 10^6$&$2.4\times 10^5$ &$2.0\times 10^5$ &$1.4\times 10^5$ &$3.4\times 10^4$ &$3.2\times 10^4$ &$1.0\times 10^5$  \\ \hline
Nov 7,  & $0$ & $3.9\times10^5$   &$1.5\times 10^5$  &$2.6\times 10^5$ &$3.8\times 10^4$ &$6.3\times 10^4$ &$2.1\times 10^4$ &$1.6\times 10^4$ &$1.0\times 10^4$ \\
2020  & $10^3$ & $1.02\times 10^8$   &$1.7\times 10^7$ &$4.0\times 10^6$ & $3.0\times 10^6$&$7.8\times 10^5$ &$2.8\times 10^5$ &$2.8\times 10^5$ &$1.4\times 10^6$ \\
(180 days) & $5\times 10^3$  & $7.2\times 10^7$   &$1.78\times 10^6$ &$2.6\times 10^5$ &$2.2\times 10^5$ &$1.7\times 10^5$ &$3.7\times 10^4$ &$7.0\times 10^4$ &$9.9\times 10^5$ \\ \hline
May 7, & $0$ & $7.8\times 10^5$ &$2.8\times 10^5$ & $6.8\times 10^5$ &$1.2\times 10^5$ &$2.7\times 10^5$ &$3.9\times 10^4$ &$3.6\times 10^4$ &$1.9\times 10^4$ \\
2021 & $10^3$ &$1.06\times 10^8$& $1.7\times 10^7$&$4.1\times 10^6$ &$3.1\times 10^6$ &$8.5\times 10^5$ &$2.9\times 10^5$ &$4.3\times 10^5$ &$3.2\times 10^6$ \\
(365 days) & $5\times 10^3$ & $7.3\times 10^7$ &$1.8\times10^6$ &$2.7\times 10^5$ &$2.3\times 10^5$ &$1.9\times 10^5$ &$4.6\times 10^4$ &$7.8\times 10^4$ &$1.2\times 10^6$  \\
\hline 
\end{tabular}
}
\caption{Prediction for states when the number of testing grows linearly at each district, based on the training data up to May 7, 2020 and the validation data period is May 8 - 14, 2020.}
\label{table:modelprediction}
\end{center}
\end{table}

\begin{table}[t]
\begin{center}
\resizebox{\textwidth}{!}{ 
\begin{tabular}{ |c|c|c |c|c|c|c|c |c|c| } 
\hline
Day & $r_2$ & MH &TN&GUJ&RAJ&MP&UP&WB&AP\\ \hline \hline
July 7,  & & & & & & & & &  \\
2020  & $0.1$ & $1.7\times 10^6$ &$1.5\times 10^6$ &$8.5\times 10^4$ &$7.4\times 10^4$ &$2.3\times 10^4$ &$8.4\times 10^4$ &$1.9\times 10^4$ &$8.1\times 10^3$ \\
(60 days) & $0.4$ & $9.3\times 10^5$ & $1.4\times 10^6$&$4.4\times 10^5$ &$5.7\times 10^5$ &$3.6\times 10^5$ &$3.3\times 10^5$ &$3.1\times 10^4$ &$6.1\times 10^4$  \\ \hline
Nov 7, & & &  & & & & & & \\
2020 & $0.1$ & $1.66\times 10^7$   &$2.4\times 10^7$ &$6.7\times 10^6$ & $5.8\times 10^6$&$8.5\times 10^5$ &$4.7\times 10^6$ &$3.3\times 10^5$ &$2.3\times 10^5$ \\
(180 days) & $0.4$  & $1.3\times 10^6$   &$1.5\times 10^6$ &$6.1\times 10^5$ &$7.5\times 10^5$ &$6.9\times 10^5$ &$3.6\times 10^5$ &$1.1\times 10^5$ &$5.1\times 10^5$ \\ \hline
May 7, & & & & & & & & & \\
2021 & $0.1$  &$1.9\times 10^7$& $2.6\times 10^7$&$8.3\times 10^6$ &$7.2\times 10^6$ &$2.2\times 10^6$ &$5.0\times 10^6$ &$1.7\times 10^6$ &$2.9\times 10^6$ \\
(365 days) & $0.4$ & $3.3\times 10^6$ &$1.6\times10^6$ &$6.6\times 10^5$ &$8.4\times 10^5$ &$8.5\times 10^5$ &$3.7\times 10^5$ &$1.8\times 10^5$ &$9.3\times 10^5$  \\
\hline 
\end{tabular}
}
\caption{Prediction for states when number of testing grows log-linearly at each district based on the training data up to May 7, 2020 and the validation data period is May 8 - 14, 2020.}
\label{table:modelprediction1}
\end{center}
\end{table}

\subsection{Statewise prediction}
We consider $8$ states which have highest number of tested positive cases. For each state, we learn a model and do the prediction of probable tested positive cases after $7$ days of the last day of validation data. We consider only those states which have sufficient data to train the model (at-least 2000 tested positive cases on May 7, 2020). Learned values of model parameters corresponding to each state are given in Table~\ref{table:modelpara}.  


After training the models corresponding to data of each state, we do the prediction of total tested positive cases in all the states on May 20, 2020. Predicted values and actual values are noted in Table~\ref{table:modelprediction}.
 

In all the experiments performed in this work, we set $\epsilon=2, \alpha=4$, and $\lambda_3$ and $\theta$ are selected using grid search and values are given in Table \ref{table:modelpara}. 


\subsubsection{Prediction with the number  of testing  as of May 7, 2020} 

Maharashtra is the most affected state in India which has at-least $30\%$ of total tested positive cases in the country on May 20, 2020. Apart from Maharashtra (MH), we consider Tamil Nadu (TN), Gujarat (GUJ), Rajasthan (RAJ), Madhya Pradesh (MP), Uttar Pradesh (UP), West Bengal (WB), and Andhra Pradesh (AP).
Maharashra has $39297$ tested positive cases as of May 20, 2020 and the proposed model predicts it as 38724 which is fairly close enough to observed value. Similarly, TamilNadu, Gujarat, Rajasthan, Madhya Pradesh (MP), Uttar Pradesh (UP), West Bengal (WB), and Andhra Pradesh (AP) have total number of tested COVID-19 infected cases as 13191, 12539, 6011, 5735, 5175, 3103, and 2560, and respective predicted values by the proposed model are 15092, 12453, 5197, 5389, 5089, 3039, and 2509 as of May 20, 2020. Thus, we conclude that the model is able to learn and predict the total number of tested positive cases in each of these states. 

In Figures~\ref{fig:maharashtra} and \ref{fig:MP}, data and corresponding curve fitting is shown in which blue dots  correspond to real data which are used as training data set, and the sky blue dots correspond to the validation data-set. The grey circles represent the trained and predicted values due to the proposed model which are following the real data very well. Grey square in plots (X=78 marked point in plots) indicates predicted value on May 20, 2020. 

    
Apart from next $7$ days of prediction, we do a prediction for after 60 days, 180 days, and 365 days from the date of validation (May 14, 2020) under different values of testing rates $r_1$ and $r_2$. Predicted values are tabulated in Tables~\ref{table:modelprediction} and \ref{table:modelprediction1} for different states and in Table \ref{table:modelprediction2} for India. Note that $r_1=0$ means when the testing statistics remain same as per testing data on May 07, 2020. However, if the number of total testing increases linearly or log-linearly as defined by Eqs.~(\ref{testing1}) and (\ref{testing}) then the number of people detected with COVID-19 increase significantly. This will be  discussed  in the  next subsection. 

The total estimate for the total number of COVID-19 infected people in India would be approximately 4,60,000 on July 7, 2020; 19,00,000 on November 7, 2020; and 46,00,000 on May 7, 2021, if the number of testing maintains the present statistics including lockdown condition.

    
\begin{figure}[H]
\begin{center}
\begin{subfigure}{.4\textwidth}
  \centering
  \includegraphics[width=1\linewidth]{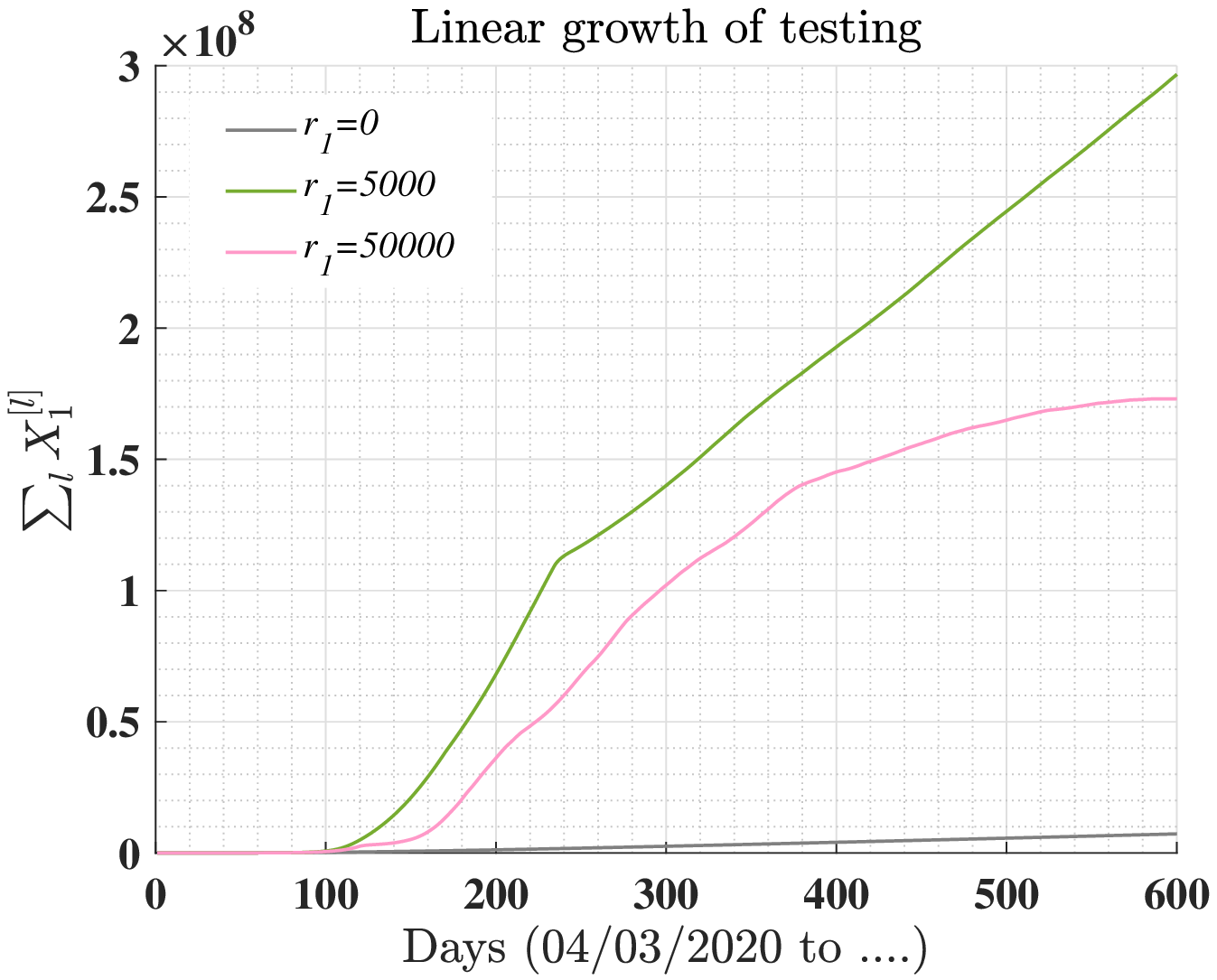}  
  \caption{}
  \label{fig1}
\end{subfigure}
\begin{subfigure}{.4\textwidth}
  \centering
  \includegraphics[width=1\linewidth]{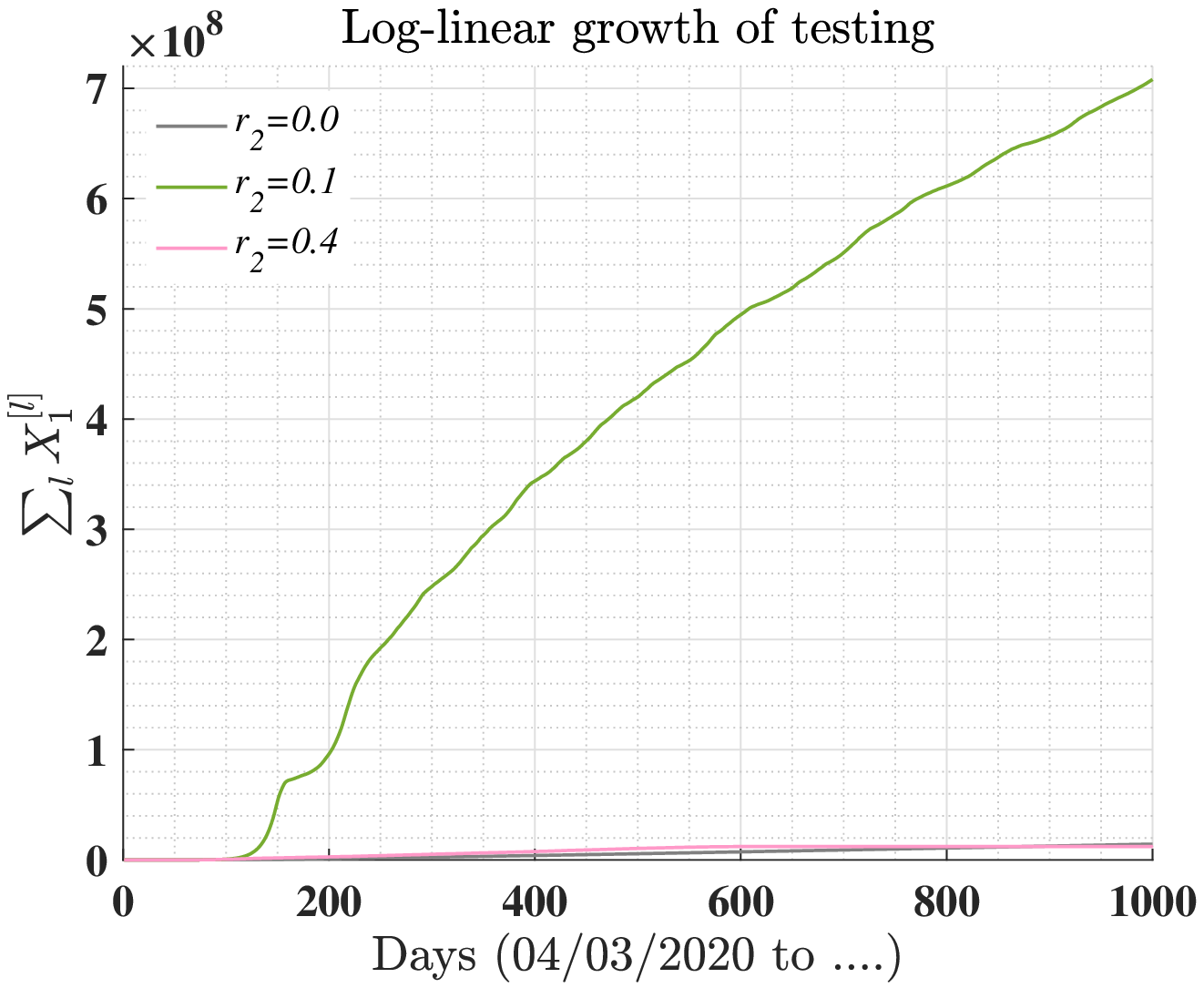}  
  \caption{}
  \label{fig16}
\end{subfigure}
\end{center}
\caption{In this figure, we show the effect of testing rate in the expected number of tested positive cases. In figures, horizontal axis represents the number of days and vertical axis represents the total number of tested positive cases ($\sum_lX_1^{[l]}$). Gain in number of testing samples is considered in two ways: (\textbf{a}) linear defined by Eq.~(\ref{testing1}), and (\textbf{b}) non-linear defined by Eq.~(\ref{testing}). In simulations, parameters are considered at state level.}
\label{fig:testingeffect}
\end{figure}

\begin{figure}[ht]
\begin{center}
\begin{subfigure}{.4\textwidth}
  \centering
  \includegraphics[width=1\linewidth]{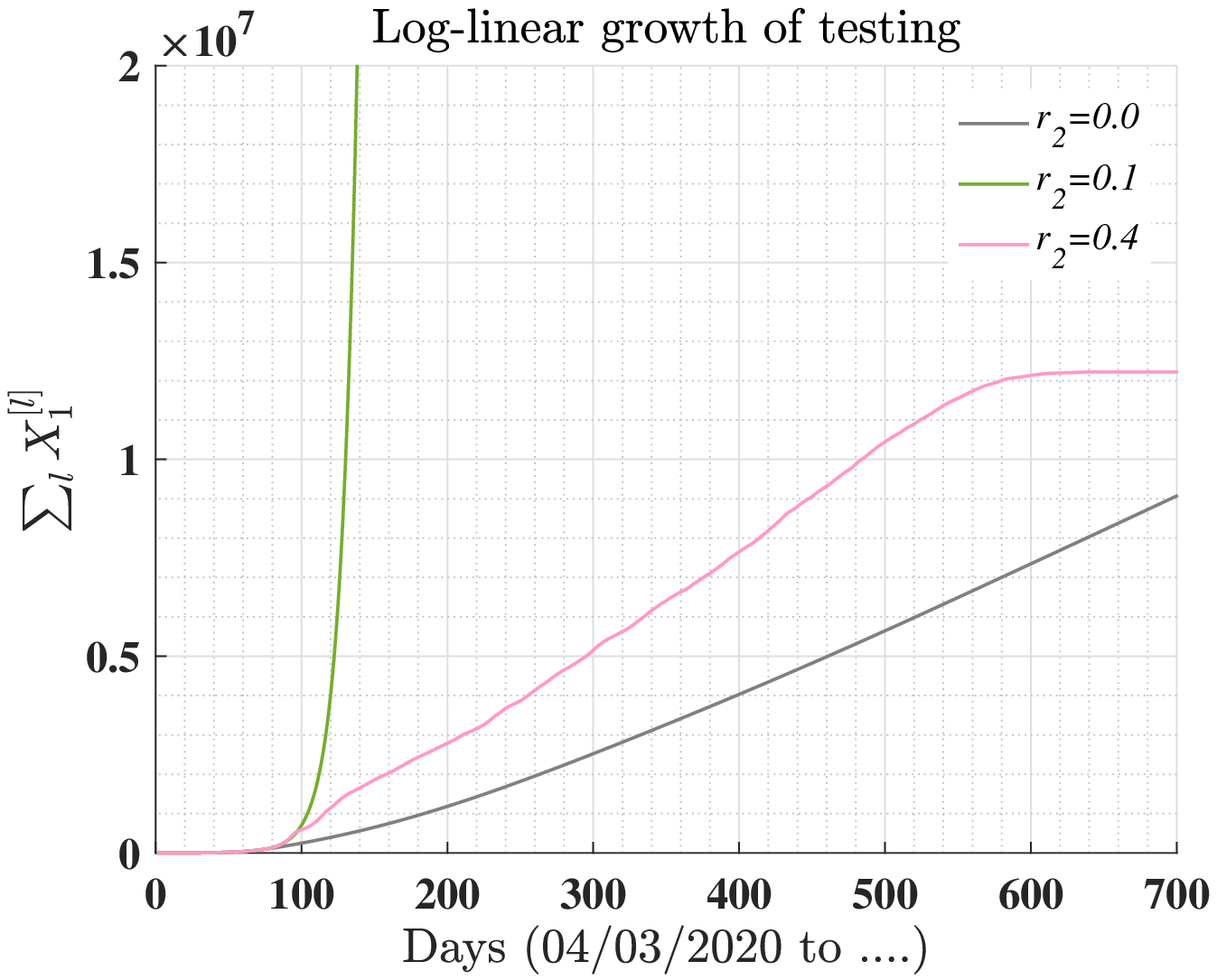}  
  \caption{}
  \label{fig1e}
\end{subfigure}
\begin{subfigure}{.4\textwidth}
  \centering
  \includegraphics[width=1\linewidth]{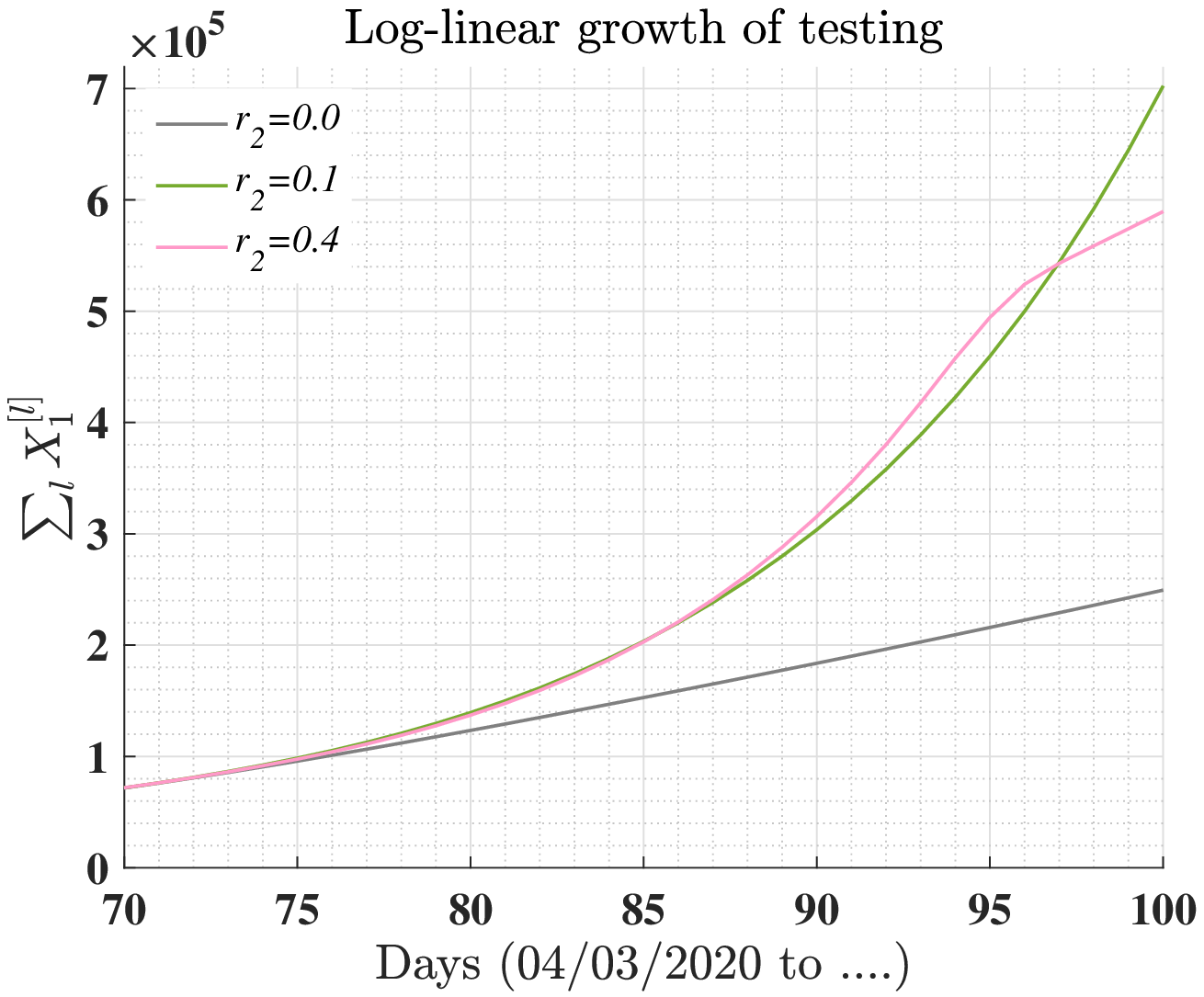}  
  \caption{}
  \label{fig16e}
\end{subfigure}
\end{center}
\caption{(\textbf{a}) Magnified view of plots in Figure~\ref{fig:testingeffect}(b). When rate of testing increases under Eq.~(\ref{testing}) with $r_2=0.4$, then spreading of COVID-19 under considered assumptions can be controlled. (\textbf{b}) In this figure, we discuss that how under different testing rates, number of positive cases get changes and testing is able to trace all the branches of infection spreading after certain time limit.  In simulations, parameters are considered at state level.}
\label{fig:testing1}
\end{figure}


\subsubsection{Effect on the total number of infected  with linear and log-linear increment in  testing} 

Currently, in India the number of testing conducted per day is approximately $1,00,000$ samples per day. However, training the proposed model over the real data provides the rate of testing $r_2=0.28.$ The testing data is not available district level but at the state level. After learning the data, the model uses different rates of gain in testing for validation. If $r_2=0$ then it means that per day testing is constant for the entire period of interest. In order to observe the effect of statistics of testing on the count of total number of people infected with COVID-19, we propose two types of growth in testing: linear and log-linear. 

First we perform this experiment when testing is increased linearly using Eq.~(\ref{testing1}), for $r_1=5000$ and $r_1=50000$. As follows form Figure~\ref{fig:testingeffect}, we consider three cases: (I) Per day testing is constant and corresponding curve is in grey color which is lowest among all three. The number of total cases of COVID-19 increasing slowly due to less number of testing over a huge population. (II) Increment of testings per day with $r_1=5000$ results into a drastic change in number of infected cases and it is increasing continuously and after certain time the rate of infection goes down but does not contain. This implies that this rate of testing may not control the spread of the disease.  (III) Finally, When the increment of testing grows linearly with $r_1=50000$ then the simulation shows that the number of infected cases  stabilizes and the spread gets contained. 
During the simulation, we consider parameter values at state level.

    
Now we discuss the effect of testing when it is increasing at log-linear rate under Eq.~(\ref{testing}). We consider the following cases: (I) The number of testing is constant with the value as of May 7, 2020. (II) If the gain in the rate of testing is $r_2=0.1$ then it does not stabilize  (III) When $r_2=0.4$ the simulation result shows stabilizes the spread and the spread can be controlled. Number of infected cases get stabilized after certain time limit, a magnified view is shown in Figure~\ref{fig:testing1}.

\begin{figure}[H]
\begin{center}
\begin{subfigure}{.45\textwidth}
  \centering
  \includegraphics[width=1\linewidth]{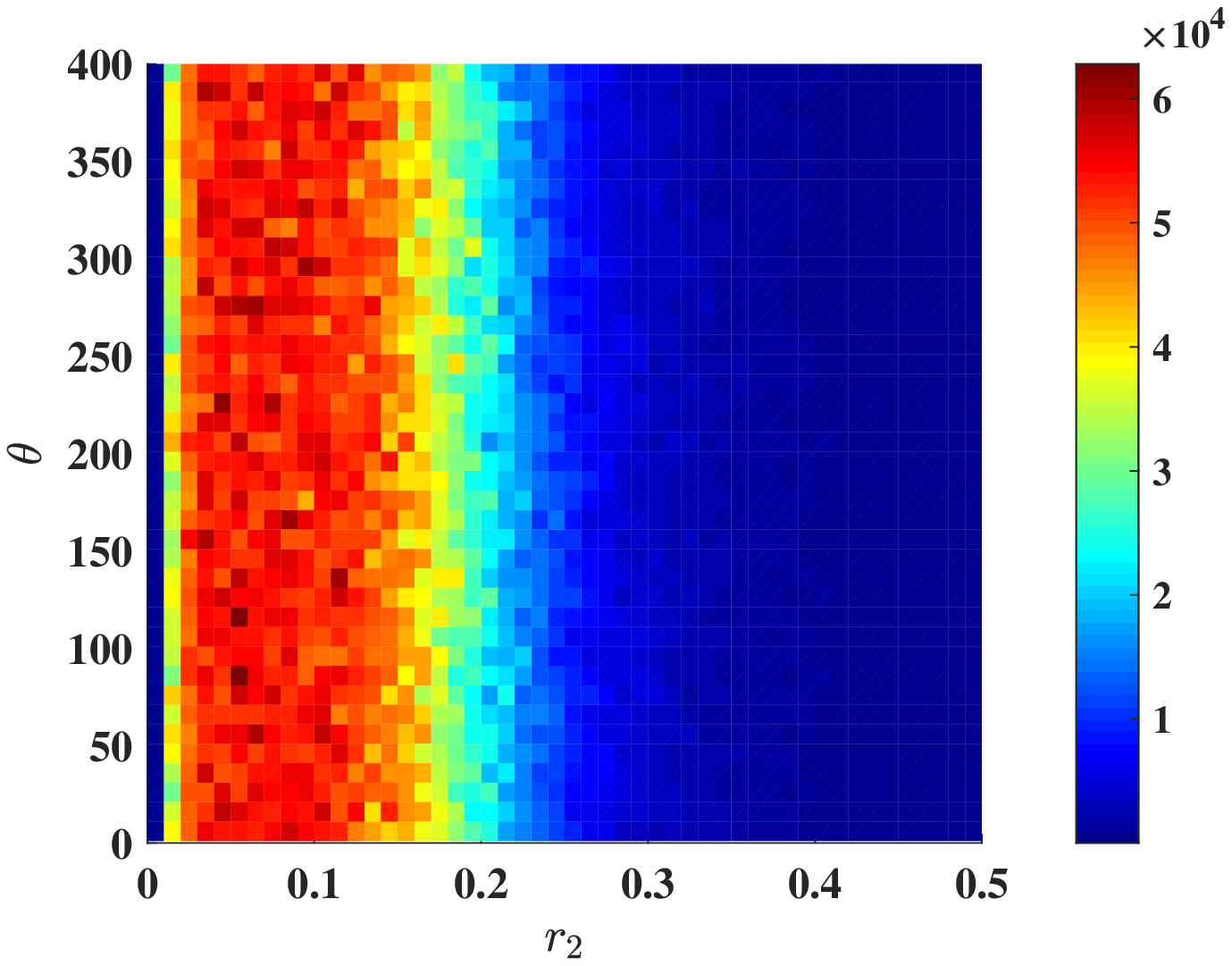} 
  \caption{$\lambda_3(t)=1/100$}
  \label{fig1g}
\end{subfigure}
\begin{subfigure}{.45\textwidth}
  \centering
  \includegraphics[width=1\linewidth]{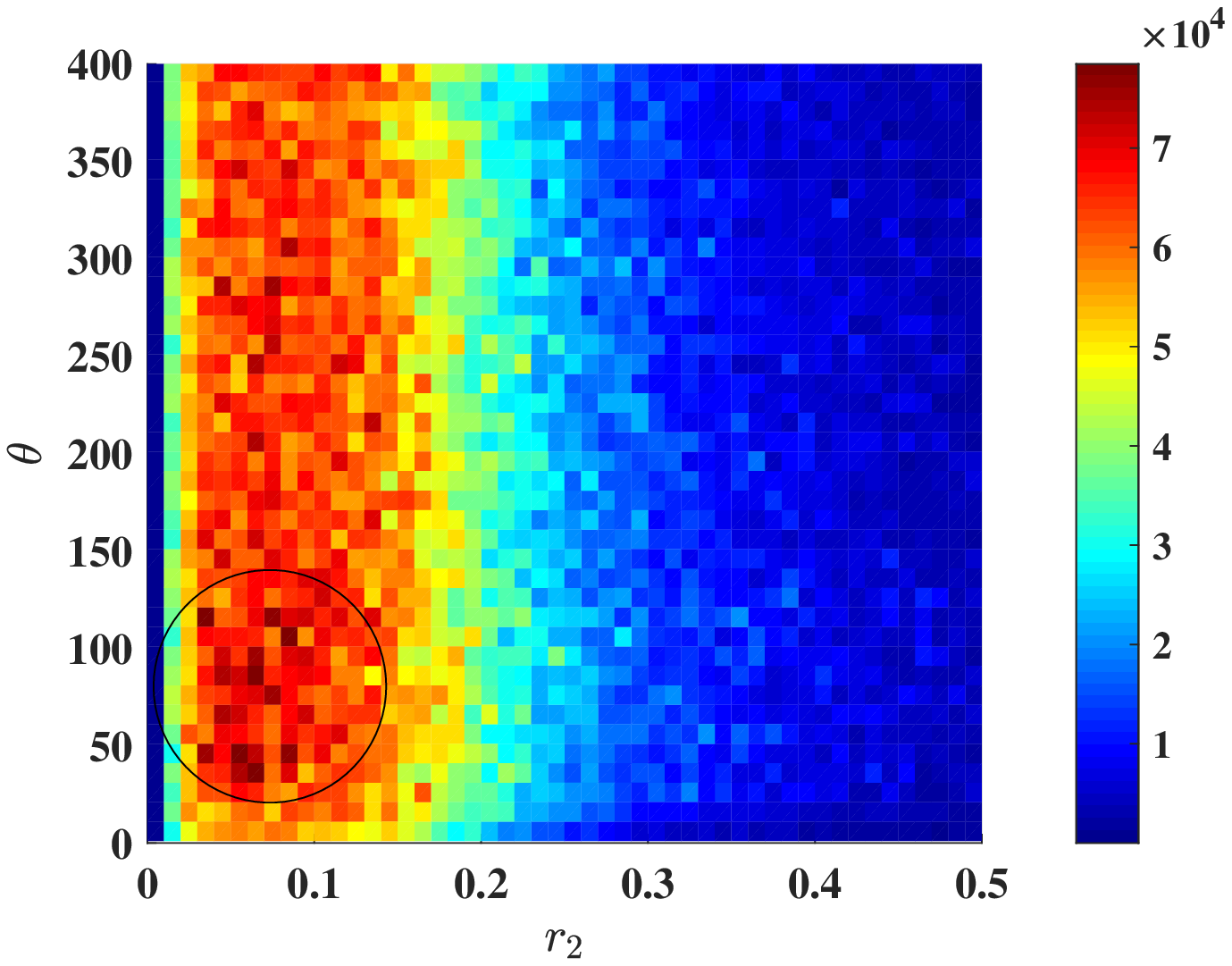}
  \caption{$\lambda_3(t)=1/10$}
  \label{fig16g}
\end{subfigure}
\end{center}
\caption{The plots shown the effect of mobility parameter $\theta$, $\lambda_3^{[l]}(t)$, and testing $r_2$. There three regions in both the sub-figures: (I) left narrow region corresponds to less number of tested positive cases as compared to middle region (II) which is followed by wide spread region (III).
It signifies that after certain rate of testing $r_2$, infection spreading can be controlled before its pandemic like situation. (\textbf{a}) when $\lambda_3^{[l]}(t)(=1/100)$ is very less then $\theta$ does not show its impact. Dark red patches are corresponding to points ($r_2,\;\theta$) which has large number of tested positive cases and these points are scattered all over the middle (II) region (highest value of tested positive cases is almost $6\times10^6$). While in (\textbf{b}) $\lambda_3^{[l]}(t)(=1/10)$ is significantly large and it shows the contribution of local mobility in spreading of infection. In plot, it is observe that dark red patches have more concentration inside circle (lower values of $\theta$ and $r_2$). Here in middle (red region) region, lower values of $\theta$ gives more weights to near by locations and less weights to locations at distances while higher values of $\theta$ give almost equal weights to all regions for mobility.}
\label{fig:thetaVsR}
\end{figure}

\subsubsection{Effect of the mobility parameter:} 

Now we study the effect of mobility of individuals ($\theta$),  probability of migration across locations ($\lam_3^{[l]}(t)$) in the spread of the disease. Besides, we analyze the joint effect of the parameters $\theta$, $\lambda_3^{[l]}(t)$) and the increment of number of testing $r_2$ (see Eq.~s (\ref{eqn:mod1}) - (\ref{eqn:mod4})). For instance, during lockdown, the value of $\theta$ can be considered as a small value. Indeed, the mobility  parameter takes a smaller value as compared to the distance between districts which are far apart but belong to the same state, and it is also true for states within the country during the lockdown period. Besides, the probability of migration is very less corresponds to lockdown effect. In simulation, we notice that the mobility parameter reaches the value up to 70 and the migration probability is less than 0.1.  


In Figure~\ref{fig:thetaVsR},  (in both the plots) we identify three regions: (I) The region in the middle, where number of tested positive (simulated) is highest, (II) A narrow region, left of (I) where the number of tested positive is less as compared to (I), and (III) stabilizing region, which is right to (I) that has less number of tested positive cases and it signifies that after certain value of $r_2$, infection spreading can be controlled.

We observe the following: (\textbf{a}) when $\lambda_3^{[l]}(t)=0.01$ no significant effect of $\theta$ in the number of infected cases is found and it simulates the spread in neighboring locations of a location which is infected (highest value of tested positive cases is almost $6\times10^6$). On the other hand, in (\textbf{b}) when $\lambda_3^{[l]}(t)=0.1$ then the contribution of local mobility into the spread of the disease is noticed. In the second plot of Figure~\ref{fig:thetaVsR}, observe that there are dark red patches (inside circles) corresponding to lower values of $\theta$ and $r_2$. In the middle (red region), the number of infected cases is more as compared to the previous case. 


Moreover, note from Eq.~ (\ref{weight}) that if the value of social mobility parameter $\theta$ is much larger  than $\max_{k,l\in\mathcal{V}(t)} d_{kl}$ then the traffic flow indicator $w_{kl}$ becomes almost uniform  for all $k$ and $l.$ On the other hand, a small value of $\theta$ generates more traffic flow between neighboring locations and induce less traffic flow across locations which are a large distant apart. If $\theta$ is considered comparatively larger value then the traffic flow are almost equal across locations. 

Besides, if $\theta$ is assigned in such a way that it induces higher traffic flow between the locations which are significantly infected then the infected people in both the corresponding location significantly increase. Consequently, the dark red patches are wide spread over the middle region in Figure~\ref{fig:thetaVsR}(a). Therefore, it is reasonable to conclude that lower values of $\theta$ corresponds to dark red patches which signifies local transmission, and larger values of $\theta$ corresponds to long-distance transmission of COVID-19.


\begin{figure}[t]
\begin{center}
\begin{subfigure}{.4\textwidth}
  \centering
  \includegraphics[width=1\linewidth]{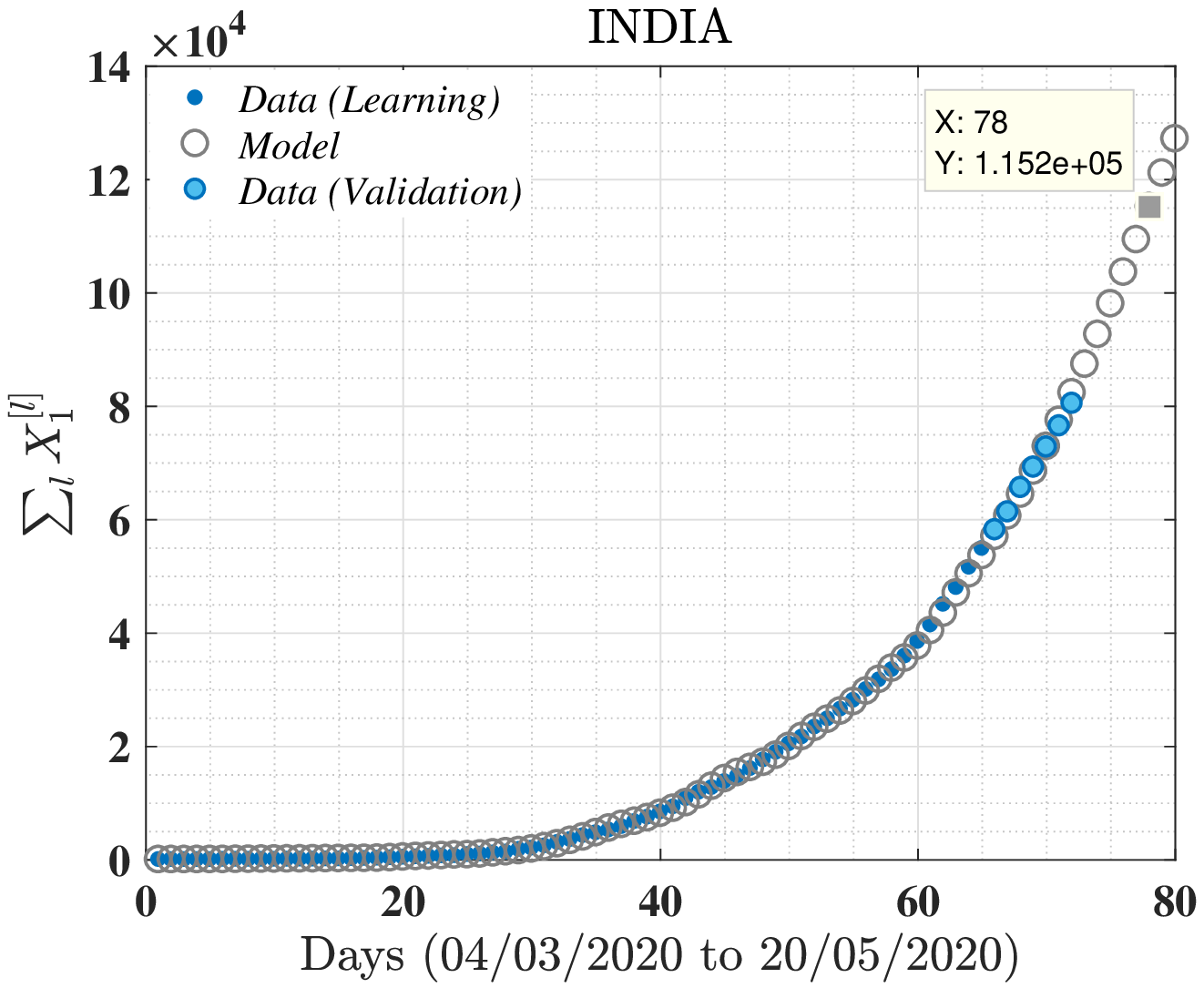}  
  \caption{}
  \label{fig1h}
\end{subfigure}
\begin{subfigure}{.4\textwidth}
  \centering
  \includegraphics[width=1\linewidth]{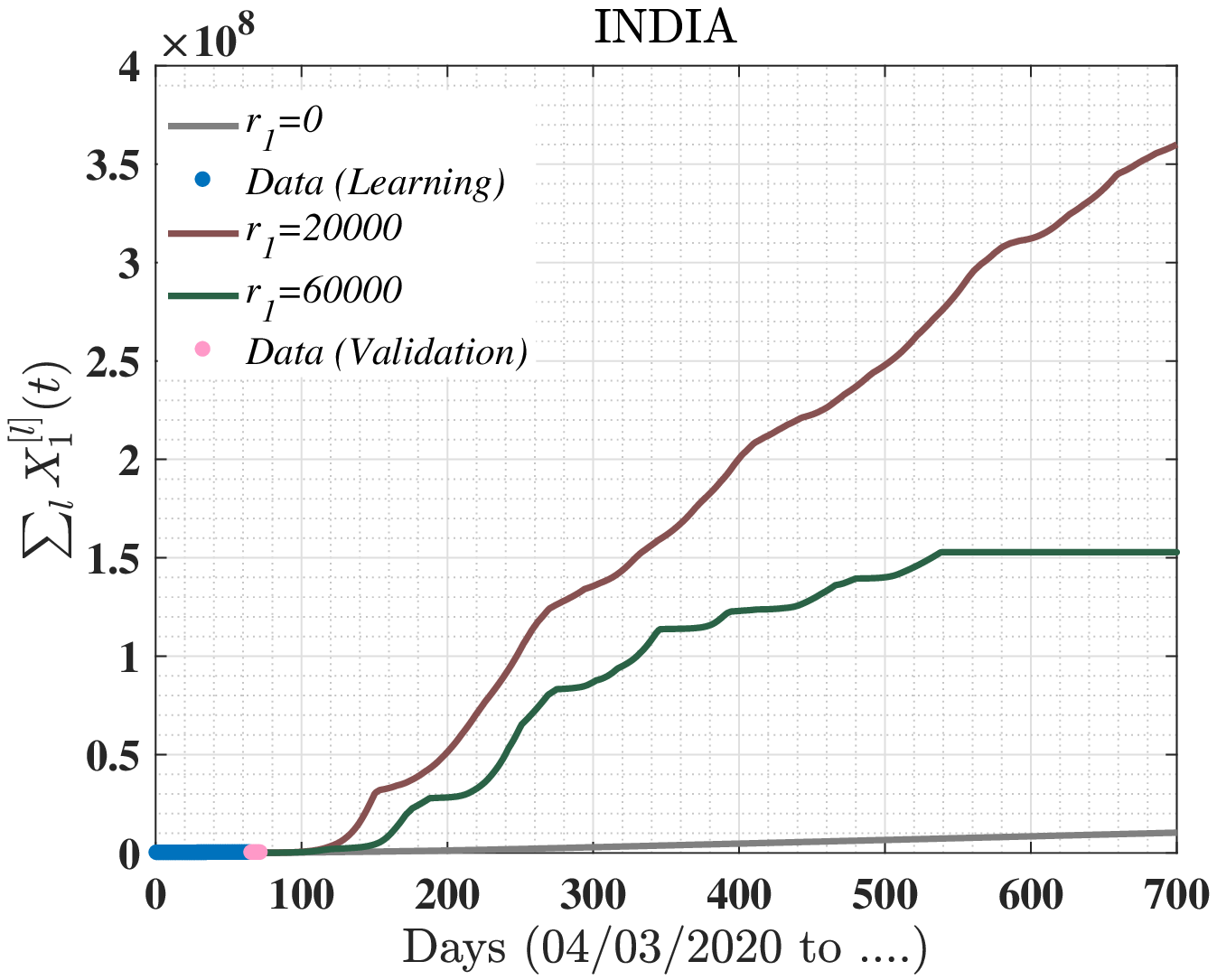}
  \caption{}
  \label{fig16h}
\end{subfigure}
\end{center}
\caption{(\textbf{a}) Blue dots are corresponding to data points used to train the model, grey circles are corresponding to trained and predicted values, sky blue dots are data points used for validation of prediction. Model is trained for infection data of 65 days, and prediction has been performed for next $7$ days with high accuracy (validation set). Sky blue dots (real data)  are close enough to corresponding predicted values (grey circles). There is shown a test point, corresponding to which, the values of tested positive cases is $1.152\times10^5$, and $1.122\times10^5$ is the actual value of it. (\textbf{b}) Cumulative number of tested positive cases under different values of testing parameter $r_1$. Plot corresponds to analysis of the status of the stabilization of  spread of COVID-19.}
\label{fig:indiaVsstate}
\end{figure}

   \begin{table}[ht]
\begin{center}
\begin{tabular}{ |c|c|c||c|c| } 
\hline
Day & $r_1$ & INDIA & $r_2$ & INDIA\\ \hline \hline
May 20, & $0$ &$1.12\times 10^5$ &  & \\
2020 & $0$ &$1.15\times 10^5$ &  & \\ \hline
July 7,  & $0$ &$4.6\times 10^5$ & & \\
2020  & $2\times 10^4$ &$5.3\times 10^6$ &$0.1$ &$6.3\times 10^6$ \\
(60 days)  & $6\times 10^4$  &$2.0\times 10^6$ & $0.4$ &$1.3\times 10^6$\\ \hline
Nov 7,  & $0$ &$1.9\times 10^6$ & & \\
2020   & $2\times 10^4$  &$8.8\times 10^7$ &  $0.1$ &$1.87\times 10^8$\\
(180 days)   & $6\times 10^4$ &$5.9\times 10^7$ & $0.4$ & $3.77\times 10^6$\\ \hline
May 7, & $0$ &$4.6\times 10^6$ & & \\
2021  & $2\times 10^4$ &$2.2\times10^8$ & $0.1$ & $3.66\times 10^8$\\
(365 days) & $6\times 10^4$  &$1.3\times 10^8$ &$0.4$ & $8.5\times 10^6$\\
\hline 
\end{tabular}
\caption{Prediction for when number of testing grows linearly and log-linearly for certain values of $r_1$ and $r_2$ based on the training data up to May 7, 2020 and the validation data period is May 8 - 14, 2020. The average error for the training dataset for the model is $37.2070$.}
\label{table:modelprediction2}
\end{center}
\end{table}

\begin{figure}[H]
\begin{center}
\begin{subfigure}{.4\textwidth}
  \centering
  \includegraphics[width=1\linewidth]{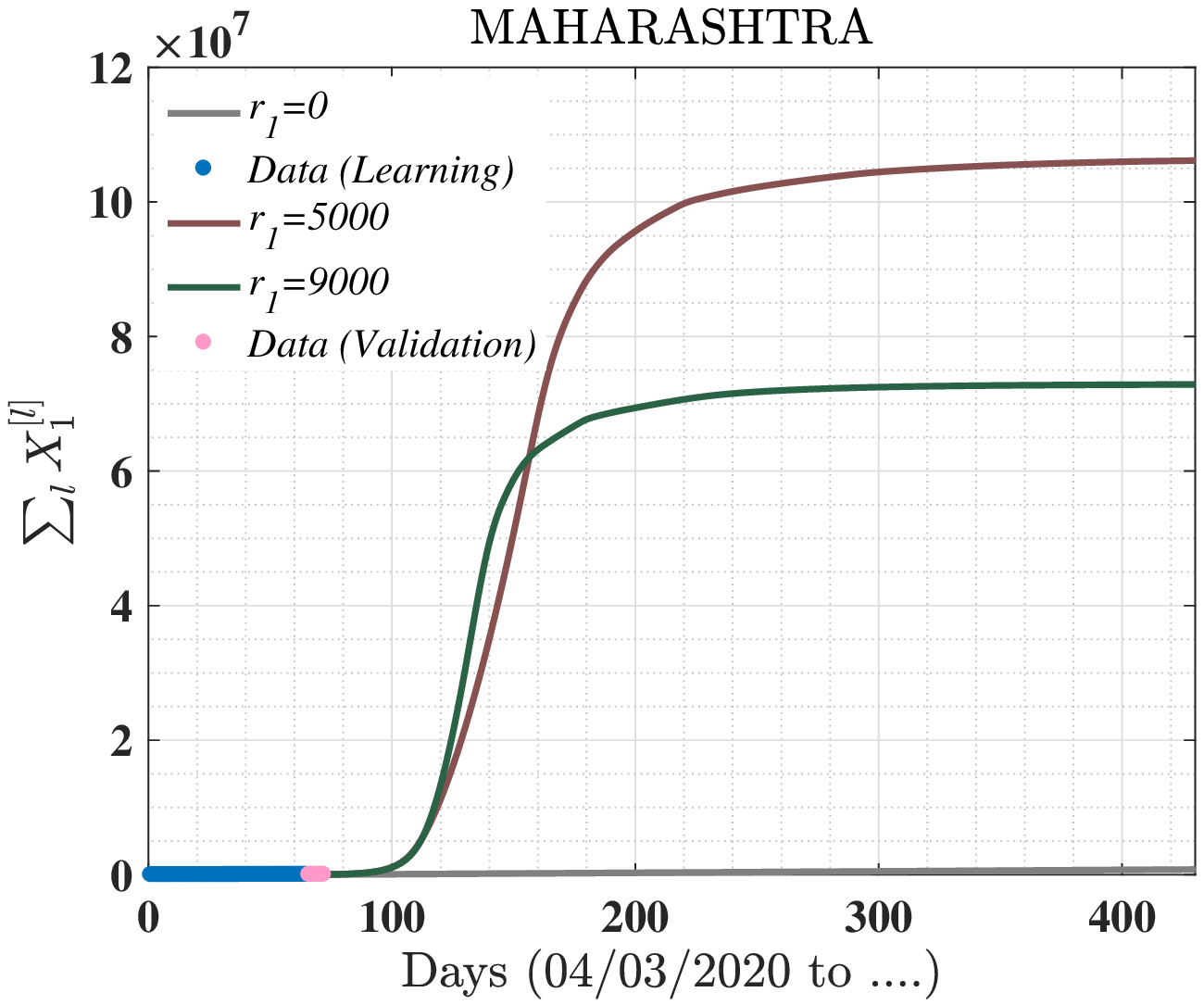}  
  \caption{}
  \label{fig1j}
\end{subfigure}
\begin{subfigure}{.4\textwidth}
  \centering
  \includegraphics[width=1\linewidth]{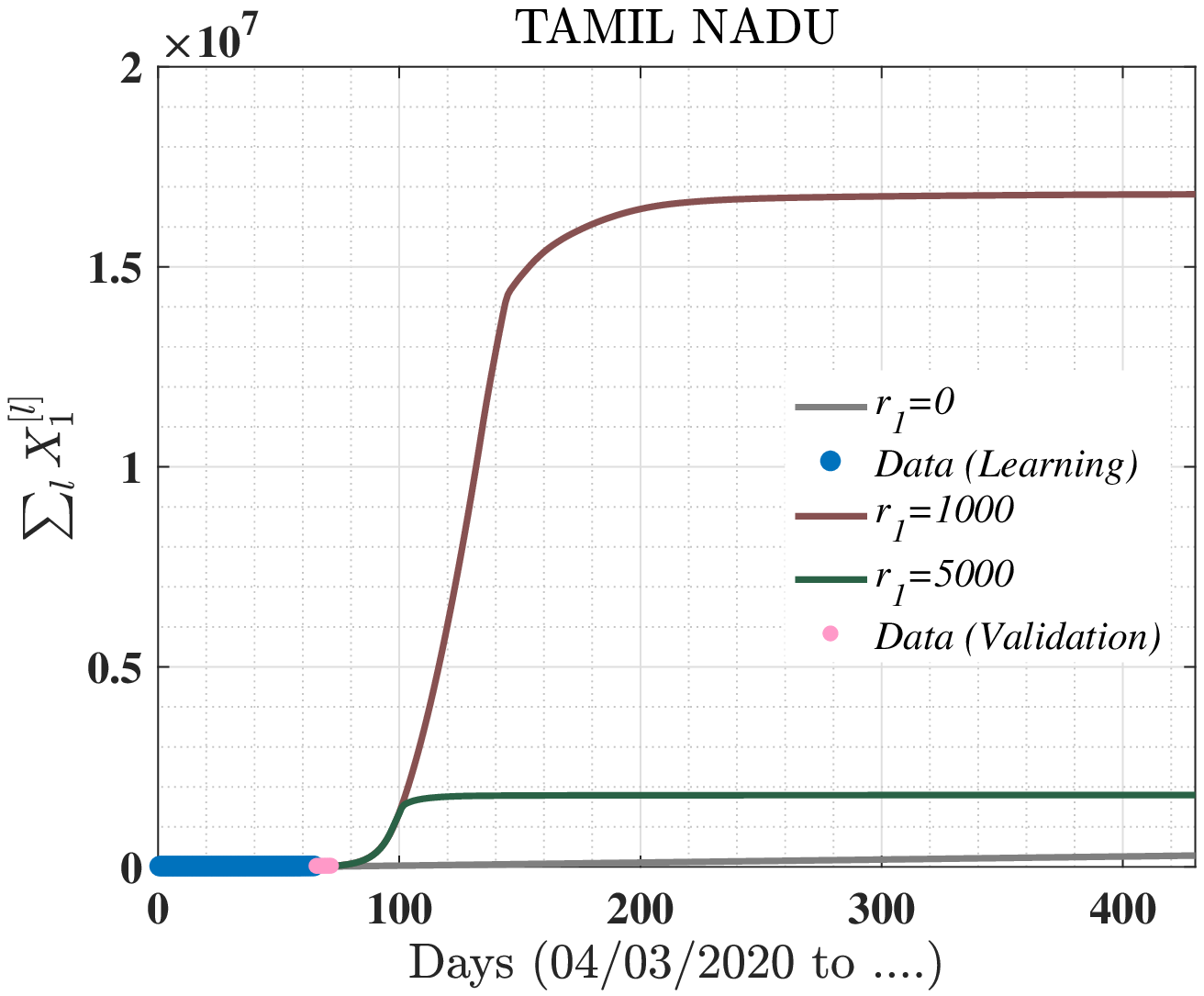}  
  \caption{}
  \label{fig16j}
\end{subfigure}

\begin{subfigure}{.4\textwidth}
  \centering
  \includegraphics[width=1\linewidth]{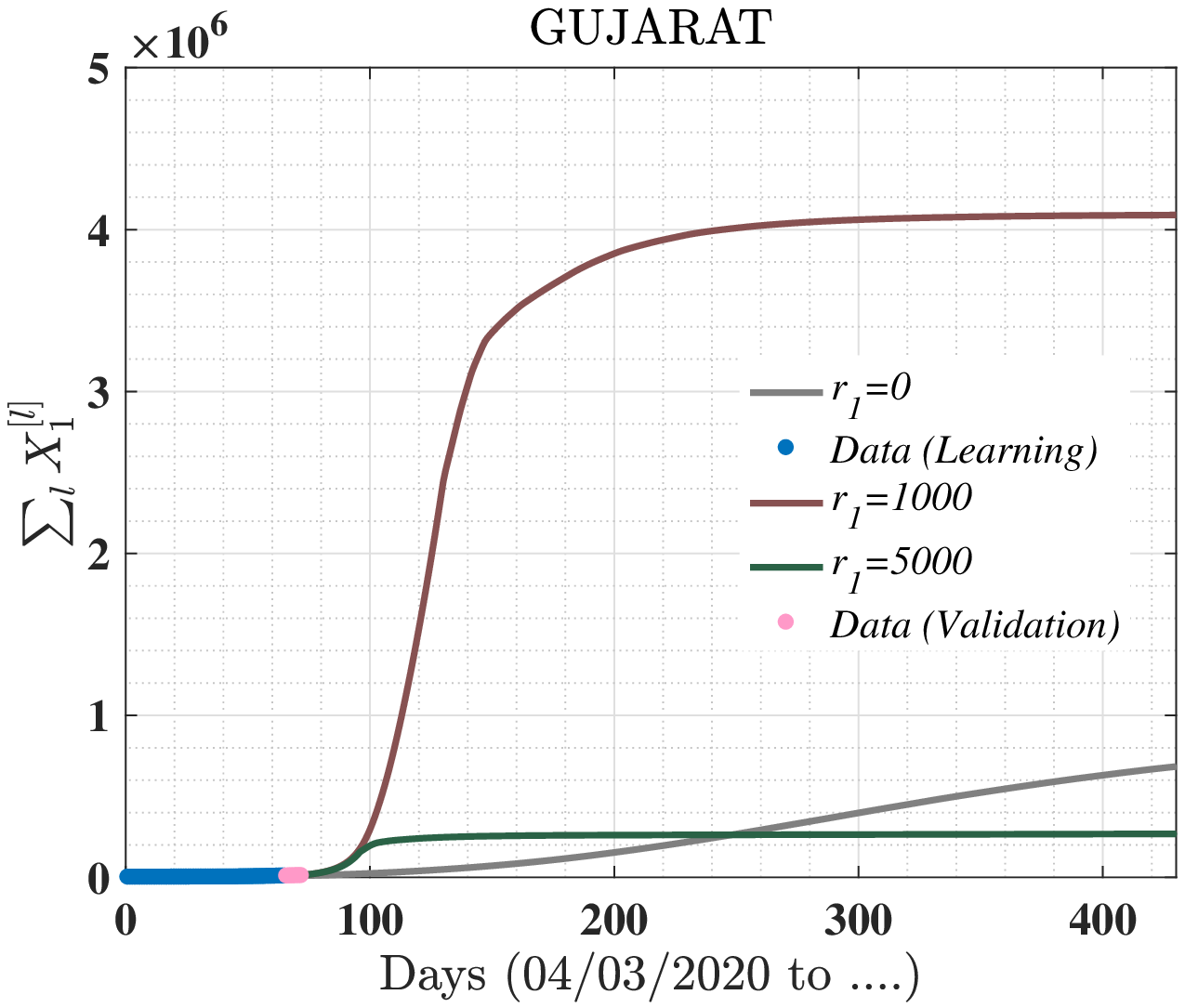}  
  \caption{}
  \label{fig1k}
\end{subfigure}
\begin{subfigure}{.4\textwidth}
  \centering
  \includegraphics[width=1\linewidth]{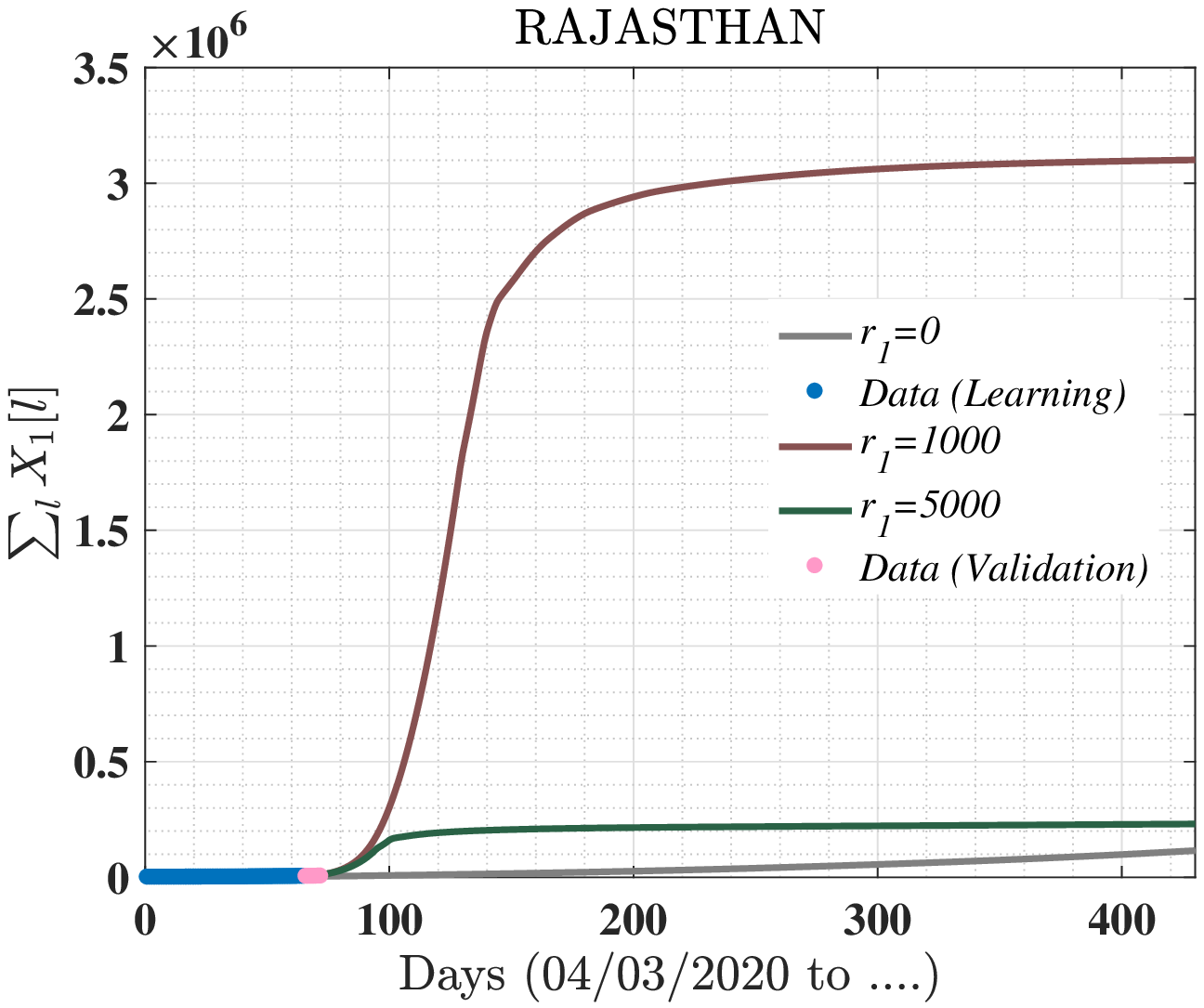}  
  \caption{}
  \label{fig16k}
\end{subfigure}
\end{center}
\caption{Cumulative number of tested positive cases under different values of testing parameter $r_1$. Plots are corresponding to analysis of the status of the stabilization of  spreade of COVID-19.}
\label{fig:stable1}
\end{figure}

\begin{figure}[h]
\begin{center}
\begin{subfigure}{.4\textwidth}
  \centering
  \includegraphics[width=1\linewidth]{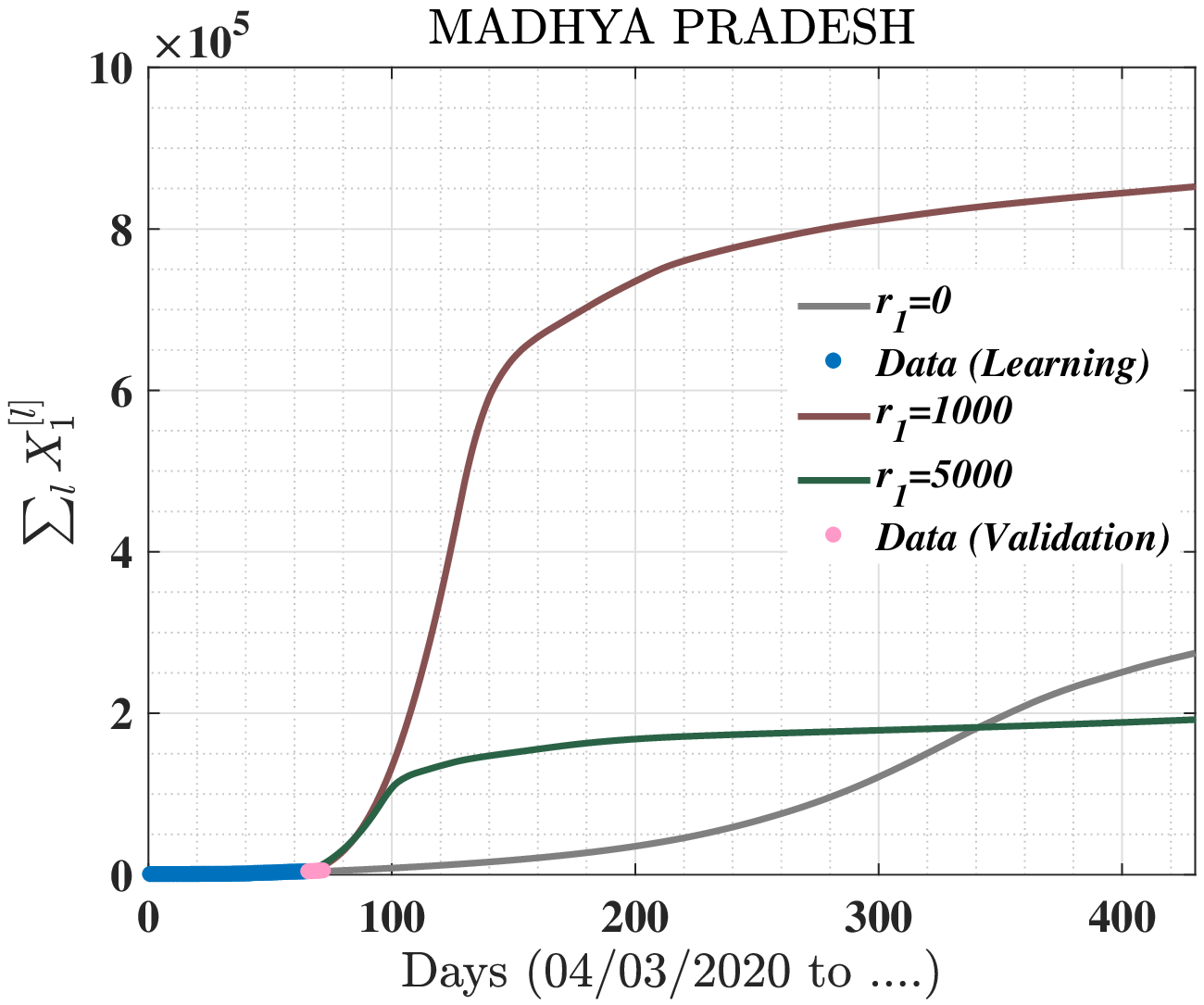}  
  \caption{}
  \label{fig1l}
\end{subfigure}
\begin{subfigure}{.4\textwidth}
  \centering
  \includegraphics[width=1\linewidth]{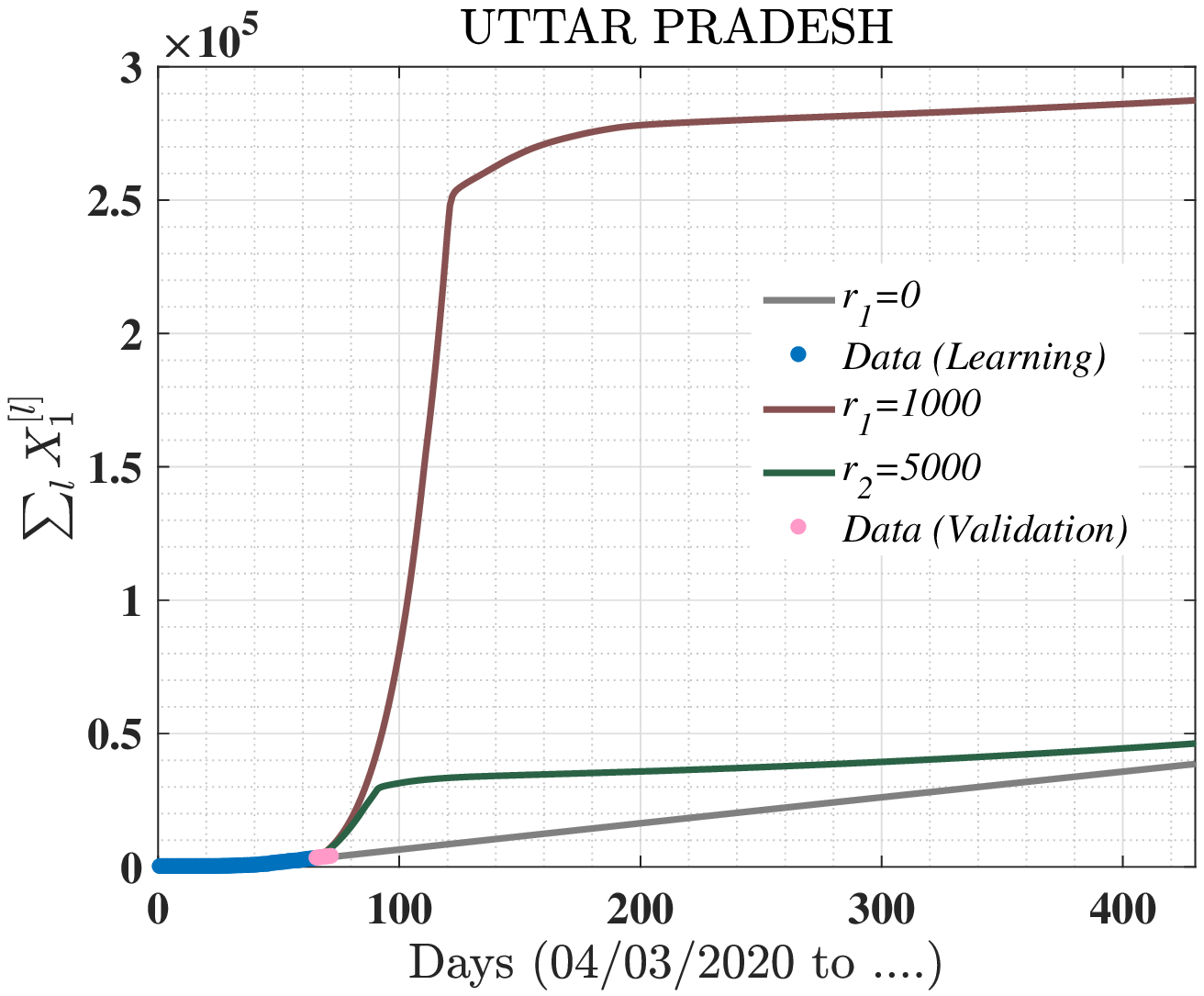}  
  \caption{}
  \label{fig16l}
\end{subfigure}

\begin{subfigure}{.4\textwidth}
  \centering
  \includegraphics[width=1\linewidth]{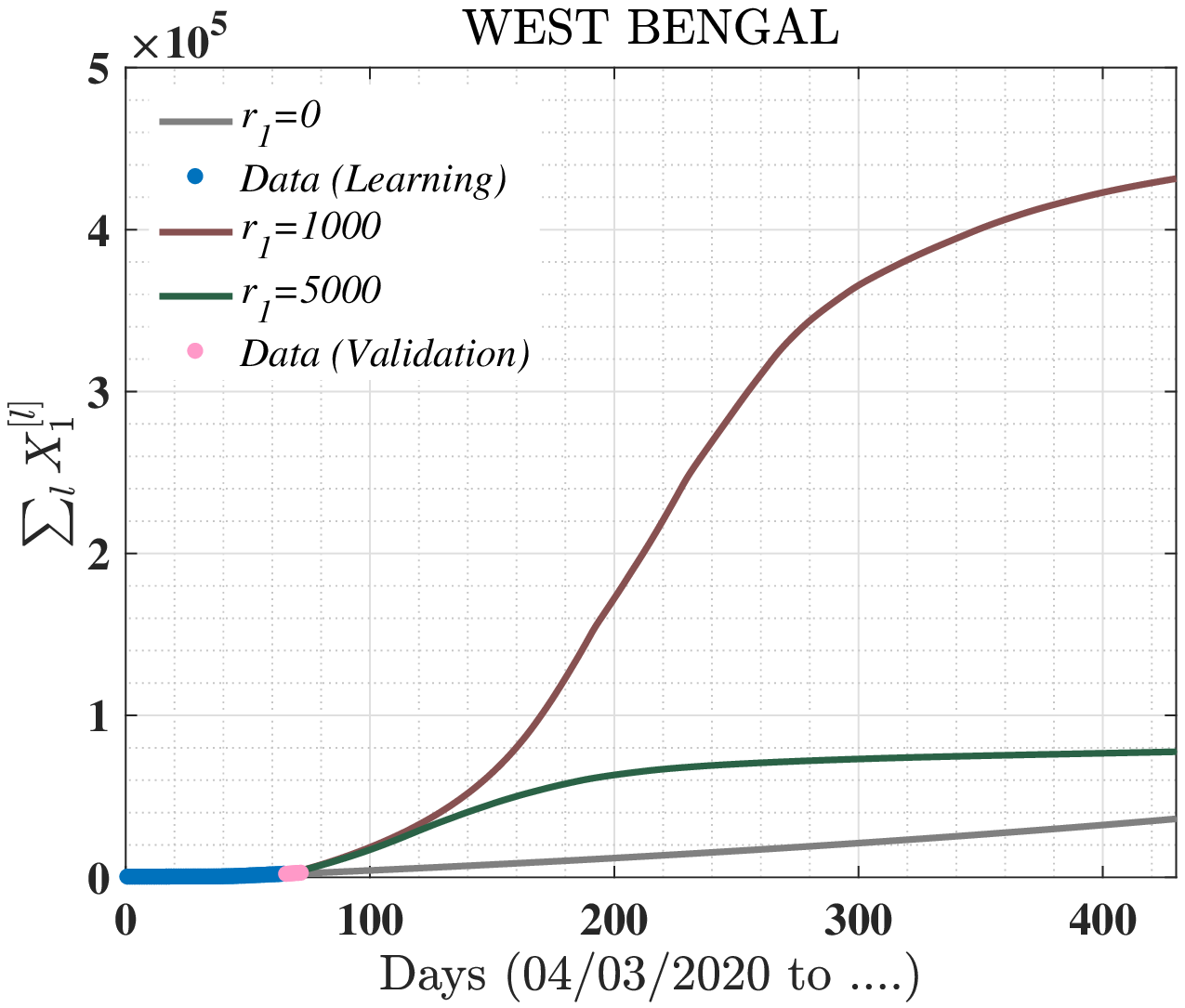}  
  \caption{}
  \label{fig1m}
\end{subfigure}
\begin{subfigure}{.4\textwidth}
  \centering
  \includegraphics[width=1\linewidth]{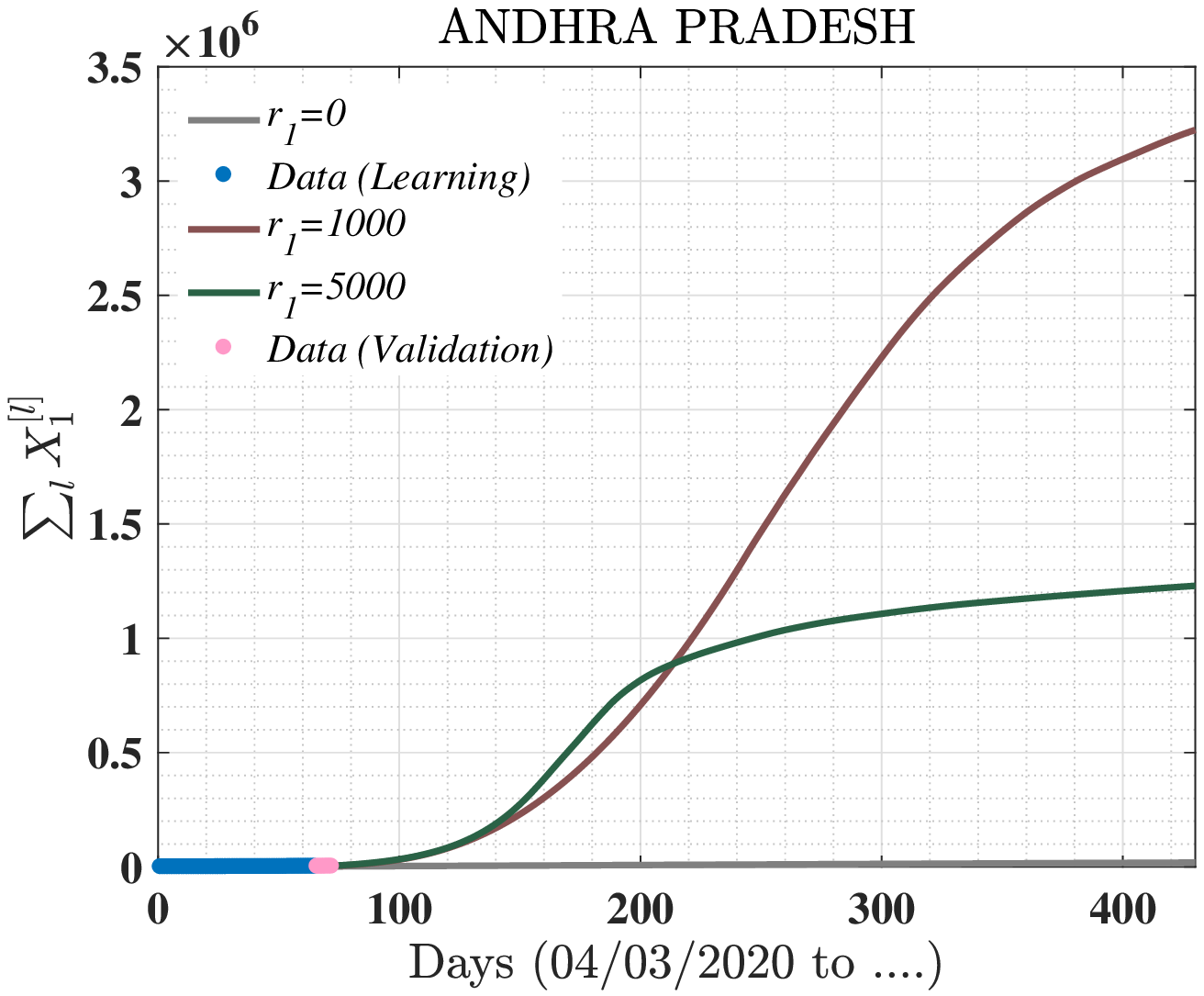}  
  \caption{}
  \label{fig16m}
\end{subfigure}
\end{center}
\caption{Cumulative number of tested positive cases under different values of testing parameter $r_1$.  Plots are corresponding to analysis of the status of the stabilization of spread of COVID-19.}
\label{fig:stable2}
\end{figure}

\subsection{Prediction for India}\label{sec:predictionIndia}

In this section, we do the prediction at country level. $X_1^{[l]}(t)$ is defined at state level and model is trained using the infection diffusion data of covid-19 spreading from March 4, 2020 to May 14, 2020. In Figure~\ref{fig:indiaVsstate}(a), blue dots are corresponding to data points used to train the model (from March 4, 2020 to May 7, 2020), grey circles are corresponding to trained and predicted values, sky blue dots are data points used for validation of prediction (from May 8, 2020 to May 14, 2020) \cite{data1}. In this plot, we have shown that model is trained for infection data of 65 days, validated using next $7$ days data, and prediction has been performed for next $7$th days almost accurately. Sky blue dots (real data) almost coincide with the centres of corresponding to the predicted values (grey circles). Here training error is $37.2070$. 


Apart from short range prediction, we also do long range prediction in which we do the prediction of possible number of tested positive cases after 60 days, 180 days, and 365 days are noted in the last column of Table~\ref{table:modelprediction2} under different testing rates $r_1$ and $r_2$. We consider $r_1=0,\;20000,\; 60000$, and $r_2=0,\;0.1,\;0.4$; after training, $r_1=0$ means testing will continue with current volume (approximately 1,00,000 per day). 

If the number of testing increases linearly with $r_1=10^3$ then the total number of people infected with COVID-19 would be approximately 5.3 Millions on July 7, 2020; 88 Millions on November 7, 2020; and 220 Millions. For linear growth with $r_1=5\times 10^3$ in testing approximates the total number of infected people in India as 2 Millions on July 7, 2020; 59 Millions on November 7, 2020; and 130 Millions, see Table \ref{table:modelprediction2}. Also the same for the log-linear increase of testing per day is given in Table \ref{table:modelprediction2}. 

\subsubsection{Stabilization of the spread of COVID-19}
In this section, we discuss about the stabilization of spreading of COVID-19 in future. This means the number of newly affected gradually decrease, and the number of total number of infected people at country level becomes almost constant.  From the analysis of the effect of mobility parameter $\theta$ and gain in testing rate $r_2$ (log-linear) or $r_1$ (linear), from Figure~\ref{fig:thetaVsR} it can be concluded that higher testing rate is more effective as we know that presently available data is obtained under very less mobility rate. However, as the mobility will increase after lifting the nationwide lockdown, the infection will presumably spread very fast. 

Here, we demonstrate time series analysis of infection spreading under different values of testing parameter $r_1$ for all the states and India. From Figures~\ref{fig:indiaVsstate}(b) and \ref {fig:testing1}(a) for country level, where as    \ref{fig:stable1}, and \ref{fig:stable2} for state level show the stabilization of  tested positive cases  with increasing   number  testing of   after certain threshold.

\begin{figure}[ht]
\begin{center}
\begin{subfigure}{.4\textwidth}
  \centering
  \includegraphics[width=1\linewidth]{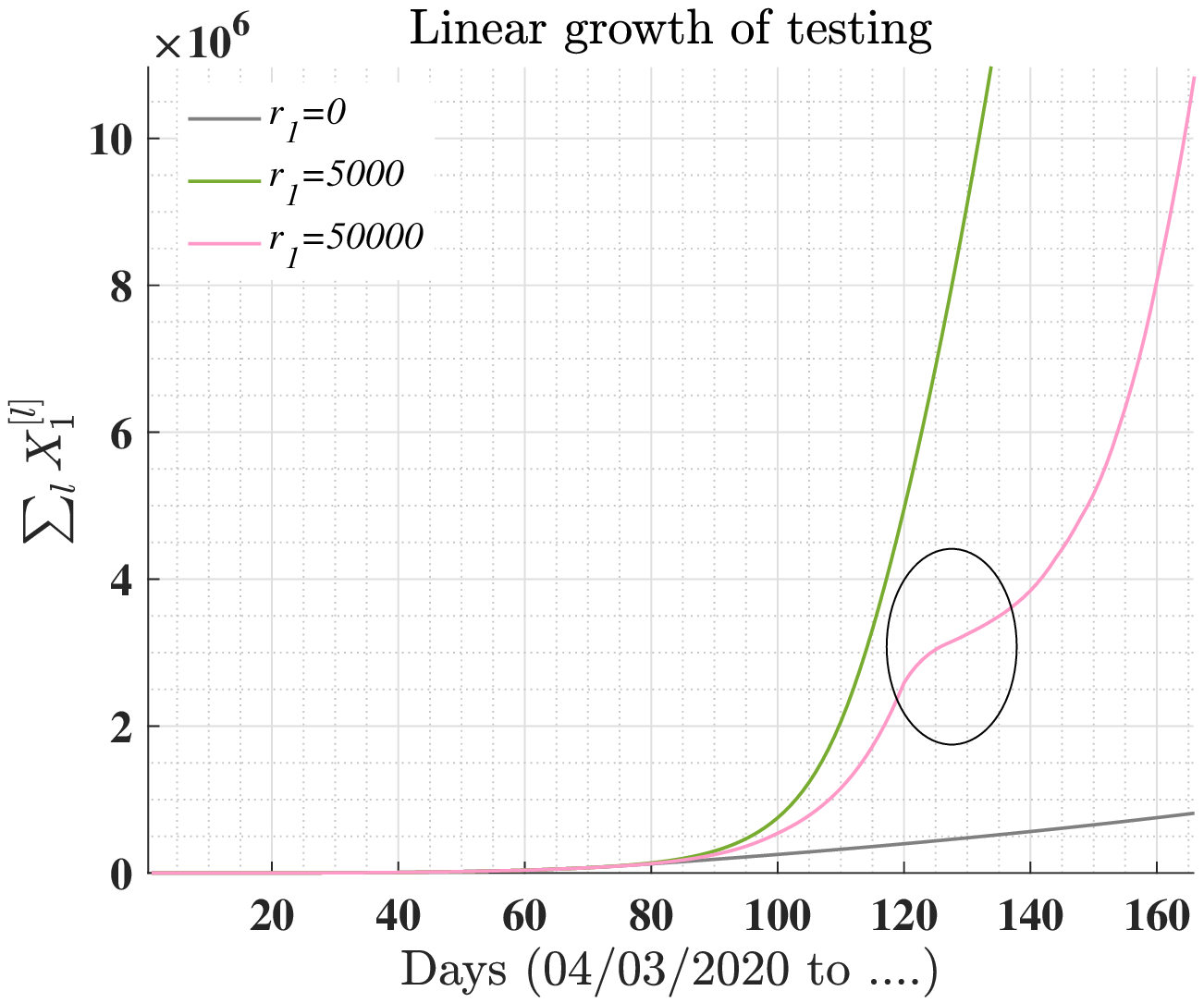}  
  \caption{}
  \label{fig1n}
\end{subfigure}
\begin{subfigure}{.4\textwidth}
  \centering
  \includegraphics[width=1\linewidth]{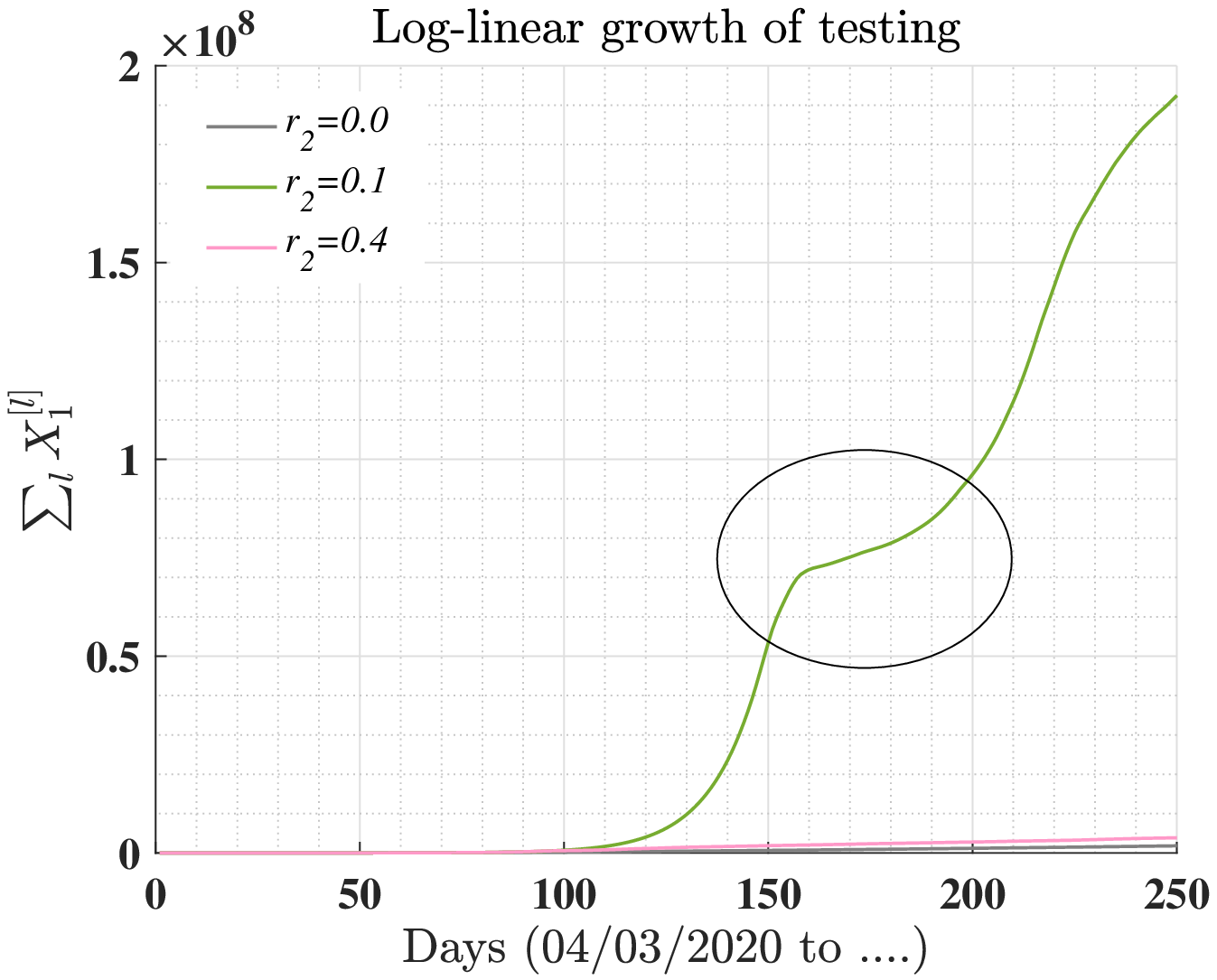}  
  \caption{}
  \label{fig16n}
\end{subfigure}
\end{center}
\caption{Possibility of second wave of the spread of COVID-19. In both the figures, we observe that we have two plots (one from each sub-figure) in which once curves of total infected  cases flattened and again rise very fast. It corresponds to the second wave of the spread of COVID-19.}
\label{fig:secondwave}
\end{figure}


\subsubsection{Second wave} 
The second wave of a pandemic is often observed in a region  when interventions are effectively applied to mitigate 
the spread of the disease and but are then lifted \cite{hulshof2020not}. In the proposed model of this paper, the second wave can be examined under the following scenarios: (I) If the number of testing performed daily is not enough, that is, it is at per with the cumulative number of all social contacts of previously detected people with COVID-19, then there would not be any sign of second wave; (II) If the number of testing is large enough such that next day available cases to be tested is decreasing continuously and spreading will get controlled soon; and (III) If the number of testing performed daily is sufficient to detect the number of cases at present state of the number of infections but somehow due to few events, (for example, if number of tested positive daily decreases and also the number of testing to be performed is below the required limit) then a second wave of diffusion may be observed. Thus it is important to track the last trail of infection diffusion completely to control it. Using simulation, we show that the second wave can be observed under different scenario which include: Number of testings is increased (\textbf{a}) linearly (\textbf{b}) log-linearly, in Figure~\ref{fig:secondwave}.

\section{Conclusion}

We have proposed an epidemiological model for the spread of COVID-19. The model is based on spread at local level which can be at the level of province, town, city or districts by combining a statistical approach and using the metapopulation network of infected locations. The model incorporates a few parameters which represent the effect of spread by asymptomatic or pre-symptomatic individuals, restricted mobility of individuals, and the testing statistics. Predictions of total number of tested with COVID-19 people are made at the level of state and the entire country, based on the data of testing as of May 7, 2020, and under linear and log-linear growth of testing statistics. Finally it is shown that the spread can be contained in very near future if linear or log-linear growth of testing is adapted.

The stabilization of infected cases primarily depends on the number of testing and the inter location transition of population or the strictness of lockdown. If the testing rate is low or moderate it may show less count  of infected cases. But there is a chance of second wave to hit-back. If the testing rate is sufficiently large and executed with proper sampling scheme then the count of positive will get stabilized much early. The proposed  epidemiological model can be applied and generalized for prediction of total number of tested with COVID-19 people at any country. Indeed, if metapopulation network is a network of countries then the prediction can be made at the world level based on the data of transmission of populations across the countries. \\\\

\noindent\textbf{Acknowledgement.} The authors thank Vaidik Dalal of How India Lives for his help with the data.

\bibliography{buddha_bib} 
\bibliographystyle{ieeetr}



\section{Appendix}
\begin{center}
\begin{table}
\begin{tabular}{|c|c|} 
 \hline
Date & Measures \\
 \hline \hline
January 25 &  screening for travellers from 2019-nCoV affected countries (China) \\ & at points of entry \\
\hline
February 26 & People coming from Republic of Korea, Iran and Italy or those \\ & having history of travel to these countries may be quarantined for \\ & 14 days on arrival to India\\
\hline
 March 3 & health scrrens at border crossings \\
  \hline
  March 5 & Advisory against mass gatherings\\
  \hline
  March 16 & closure of selected public institutions such as museums (incl. Taj Mahal)  \\ & until March 31 and postponement of several local elections\\
  \hline 
  March 17 & Travel of passengers from Afghanistan, Philippines, Malaysia to India \\ & is prohibited  with immediate effect. No flight shall take off from these  \\ & countries to India after 1500 hours Indian Standard Time (IST) till \\ & March 31 and will be reviewed subsequently. \\
  \hline
March 18 & ban of entry for passengers from EU countries, EFTA countries, \\ &  Turkey, UK  \\
\hline
March 22 & No international flights to take off for India from foreign airports \\ & after 0001 hrs GMT of March 22, 2020 until 0001 hrs GMT March 29, 2020. \\ & 20 hours maximum travel time. So no incoming international passengers \\ & allowed on  Indian soil (foreigner or Indian) after 2001 hrs GMT \\ & of March 22, 2020. \\
 \hline
 March 24 & Complete lockdown of entire nation for 21 days. Agriculture-Farming \\ & and allied activities exempted from Lockdown \\
 \hline
April 25  & opening of certain categories of shops. In rural areas, all shops, \\ & except those in shopping malls are allowed to open. In urban areas, \\ & all standalone shops, neighborhood shops, shops in residential complexes \\ & are allowed to open. Shops in markets/market complexes and  \\ & shopping malls are not allowed to open. It is clarified that  \\ & sale by e-commerce companies will continue to be \\ & permitted for essential goods only.\\ 
  \hline 
  May 4 & Extension of Lockdown for a further period of Two Weeks with effect \\ & from May 4, 2020 \\
  \hline
\end{tabular}
\caption{A few preventive measures taken by Govt. of India}
\label{table:measures}
\end{table}
\end{center}

\end{document}